\documentclass[10pt, conference, final, a4paper]{IEEEtran}
\ifCLASSINFOpdf

\else
\fi

\usepackage{stmaryrd}
\usepackage[utf8]{inputenc}
\usepackage{multirow}
\usepackage{multicol}
\usepackage{hyphenat}
\usepackage[hyphens]{url}
\usepackage{amsmath}
\usepackage[noend]{algpseudocode}
\usepackage{algorithmicx,algorithm}

\usepackage[pdftex]{graphicx}
\usepackage[dvipsnames]{xcolor}

\begin{document}
%
\title{FAT: An In-Memory Accelerator with \underline{F}ast \underline{A}ddition for \underline{T}ernary Weight Neural Networks}

\author{\IEEEauthorblockN{Shien Zhu\IEEEauthorrefmark{1}, Luan H.K. Duong\IEEEauthorrefmark{1}, Hui Chen\IEEEauthorrefmark{1}, Di Liu\IEEEauthorrefmark{2}, Weichen Liu\IEEEauthorrefmark{1}}
\IEEEauthorblockA{\IEEEauthorrefmark{1}School of Computer Science and Engineering, Nanyang Technological University, Singapore}
\IEEEauthorblockA{\IEEEauthorrefmark{2}HP-NTU Digital Manufacturing Corporate Lab, Nanyang Technological University, Singapore\\
Email: shien001@e.ntu.edu.sg, \{lhkduong, hui.chen, liu.di, liu\}@ntu.edu.sg }}




\maketitle

\begin{abstract}
Convolutional Neural Networks (CNNs) demonstrate excellent performance in various applications but have high computational complexity. Quantization is applied to reduce the latency and storage cost of CNNs. Among the quantization methods, Binary and Ternary Weight Networks (BWNs and TWNs) have a unique advantage over 8-bit and 4-bit quantization. They replace the multiplication operations in CNNs with additions, which are favoured on In-Memory-Computing (IMC) devices. IMC acceleration for BWNs has been widely studied. However, \textcolor{black}{though TWNs have higher accuracy and better sparsity than BWNs}, IMC acceleration for TWNs has limited research. TWNs on existing IMC devices are inefficient because the sparsity is not well utilized, and the addition operation is not efficient. 

In this paper, we propose FAT as a novel IMC accelerator for TWNs. First, we propose a Sparse Addition Control Unit, which utilizes the sparsity of TWNs to skip the null operations on zero weights. Second, we propose a fast addition scheme based on the memory Sense Amplifier to avoid the time overhead of both carry propagation and writing back the carry to memory cells. Third, we further propose a Combined-Stationary data mapping to reduce the data movement of activations and weights and increase the parallelism across memory columns. Simulation results show that for addition operations at the Sense Amplifier level, FAT achieves 2.00\(\times\) speedup, 1.22\(\times\) power efficiency and 1.22\(\times\) area efficiency compared with \textcolor{black}{a} State-Of-The-Art IMC accelerator ParaPIM. FAT achieves 10.02\(\times\) speedup and 12.19\(\times\) energy efficiency compared with ParaPIM on networks with 80\% \textcolor{black}{average sparsity.}

\end{abstract}



\setlength{\belowdisplayskip}{4pt} \setlength{\belowdisplayshortskip}{3pt}
\setlength{\abovedisplayskip}{4pt} \setlength{\abovedisplayshortskip}{3pt}

\section{Introduction}

Deep Convolutional Neural Networks (CNNs) have been widely adopted in computer vision \cite{CVPR_2020_EfficientDet}, natural language processing \cite{NIPS_2020_WAV2VEC}, robotics \cite{ICLR_2020_Robots} and many other fields. Deep CNNs have numerous parameters, and the convolution layers are both memory-intensive and computation-intensive, resulting in high storage cost and long latency when deployed on the edge. \textcolor{black}{Thus} deep CNNs need optimization and acceleration for the edge to achieve low storage overhead and high performance.  


Quantization is one of the effective CNN acceleration methods that reduce CNNs' computation complexity and storage cost. Quantization methods utilize low bitwidth numbers to represent the values of CNNs. Among the quantization methods, \textcolor{black}{Binary Weight Networks (BWNs) \cite{ECCV_2016_XNOR-Net,CVPR_2017_HWGQ,CVPR_2020_LS-BQNN} achieve high performance and extreme storage-saving} thanks to their 1-bit representation of weights. BWNs quantize the weights of CNNs into \{+1, -1\} to replace the computation-intensive multiplication operations with addition and subtraction operations for high speedup. \textcolor{black}{However, this aggressive quantization leads to lower accuracy.} \textcolor{black}{As both the accuracy and the speed matter, other CNN quantization methods,} including 8-bit \cite{CVPR_2020_UINT8, AAAI_2021_DA-INT8} and 4-bit \cite{CVPR_2019_INT4, NIPS_2020_INT4} integer quantization (INT8 and INT4) and ternary quantization \cite{ICLR_2017_TTQ, AAAI_2020_RTN, AAAI_2021_TRQ}, are proposed to do a trade-off between the speed and accuracy, as Table \ref{Tab-acc} shows. 

Ternary Weight Networks (TWNs) \cite{ICLR_2017_TTQ,AAAI_2020_RTN} make an excellent trade-off between BWNs and 32-bit Full-Precision (FP) CNNs. TWNs quantize the weights of CNNs into \{+1, 0, -1\} and brings several advantages, as shown in Table. \ref{Tab-acc}. First, TWNs have much higher accuracy than BWNs due to their higher representation capacity of weights. Second, the 2-bit representation of TWNs brings 16\(\times\) storage-saving compared with the 32-bit FP CNNs. \textcolor{black}{Third, just like BWNs, the addition has become the dominant operation in TWNs and guarantees high performance.} While 8-bit/4-bit integer quantization still uses computation-intensive multiplications in the convolution layers. Last, TWNs have zeros in the weights referred to as sparsity. This means the addition operations related to zero weights can be skipped to achieve even higher performance than BWNs. \textcolor{black}{The hardware and programming library support is essential to skipping the null operations \cite{ISCA_2017_SCNN,JCS_2019_Eyeriss-v2,HotChips_2020_NvidiaA100}}. Therefore, a hardware accelerator with fast and sparse addition is necessary to obtain the mentioned benefits of TWNs. 

\begin{table}[]
\centering
\caption{Top-1 accuracy on ImageNet of quantized ResNet-18.} 
\label{Tab-acc} 
\begin{tabular}{ | l | c | c | r | r | r |}
\hline
Method                           & Type  & Bitwidth & Operator   & Accuracy  & Sparsity \\ \hline
Original\cite{CVPR_2016_ResNet}  & FP    & 32-bit  & \(\times,+\) & 70.3\% & 10-40\%  \\ \hline
U-INT8\cite{CVPR_2020_UINT8}     & INT8  & 8-bit   & \(\times,+\) & 69.7\% &  -   \\ \hline
ULP-INT4\cite{NIPS_2020_INT4}    & INT4  & 4-bit   & \(\times,+\) & 69.0\% &  -  \\ \hline
RTN\cite{AAAI_2020_RTN}          & TWN  & 2-bit   & \(+\)       & 68.5\% & 40-90\%  \\ \hline
LS-BQNN\cite{CVPR_2020_LS-BQNN}  & BWN  & 1-bit   & \(+\)       & 66.1\% & 0\%      \\ \hline 
\end{tabular}
\end{table}


In-Memory-Computing (IMC) \cite{Nature_2020_IMC_devices} is an excellent choice in accelerating addition-centric CNNs, including BWNs and TWNs. As deep CNNs are both memory-intensive and computation-intensive, the memory access on hardware is an essential factor in CNN performance besides the computation. Unfortunately, traditional von Neumann architecture separates the memory from the processing unit, which results in limited memory bandwidth, long memory access latency, and high energy costs in transferring data between the memory and processing units. These challenges are known as the memory wall. IMC is a promising way to relieve the memory wall, utilizing resistant memories like Spin-Transfer Torque Magnetic Random-Access-Memory (STT-MRAM) to compute inside the memory \cite{DATE_2018_CIM-Spin}. Thanks to the massive data parallelism, ultra-high internal memory bandwidth, and reduced data movement, IMC brings high performance and \textcolor{black}{excellent} energy efficiency. In addition, State-Of-The-Art (SOTA) IMC devices can conduct Boolean functions and addition operations, making them very suitable for accelerating addition-centric CNNs \cite{TCS_2021_SRAM-IMC-AI, GLVLSI_2020_IMC-AI}, including BWNs and TWNs.

However, though TWNs have higher accuracy and better sparsity than BWNs, current IMC accelerators have not been optimized to fit the characteristics of TWNs. First, the sparsity of TWNs is not well utilized. The sparse weights of TWNs bring the need for compressed sparse formats to save storage \cite{SC_2019_SparseMVM, ISPASS_2021_SparseMM}. \textcolor{black}{However, the compression and decompression modules will increase the complexity of the accelerator design, the area cost, and the power. Moreover, the compressed sparse format has reduced storage benefit on TWNs because it stores the indices of the 2-bit non-zero weights in longer bit-width (e.g., 8-bit index numbers).} It is a challenge to deal with the sparsity and the sparse data format. As a result, the sparsity of TWNs has not been exploited to skip the operations of neural networks on existing IMC accelerators. For example, TiM-DNN \cite{TVLSI_2020_TiM-DNN} and XNOR-SRAM \cite{JSSC_2020_XNOR-SRAM} all process the ternary neural networks without considering the sparsity.

Also, addition operations have replaced the multiplications for higher performance in TWNs, but the addition operations in existing IMC devices are not efficient. The mainstream STT-CiM series \cite{TVLSI_2017_STT-CiM, ISLPED_2019_FEFET-CiM, TN_2020_VCS-CiM}, ParaPIM series \cite{ASPDAC_2019_ParaPIM, TCAD_2019_MRIMA}, and GraphS series \cite{DATE_2019_GraphS,NanoArc_2019_ParaPIM,ASP-DAC_2020_CA-DNN-PIM} \textcolor{black}{designs are excellent traditional IMC architectures or application-specific accelerators}, but their addition schemes are not efficient. They either need to wait for the carry propagating to the last bit or write back the carry to the memory cells and read out the carry back and forth. Applying the same addition schemes directly to processing TWNs will lead to long latency and high power consumption.

In this paper, we proposed FAT as an STT-MRAM based IMC accelerator for high-performance and energy-efficient TWN inference. First, we propose a Sparse Addition Control Unit at the architecture level, which utilizes the sparsity of TWNs to \textcolor{black}{speed up} the convolution and fully connected layers. We combine the \textcolor{black}{2-bit} weights with the control signals of the memory arrays to skip the addition operations where the weights are zeros. \textcolor{black}{This design also benefits FAT with the 16\(\times\) storage reduction without compressed sparse formats.}

Second, we propose a fast addition scheme at the circuit level based on the memory Sense Amplifier. We store the intermediate carry in a latch and compute the summation of two operands in only one step. This \textcolor{black}{addition scheme} avoids the time overhead of both waiting for the propagation of the carry \textcolor{black}{signal} and writing the carry back to the memory.  

\textcolor{black}{Third, we propose a Combined-Stationary data mapping scheme to fit the new computation pattern in FAT and fully utilize the computational capacity. The Image-to-Column based flexible mapping can reduce the data movement of activations and weights for lower energy, achieve near 100\% column para-llelism in memory arrays for higher performance, and balance the endurance of STT-MRAM cells for a longer lifetime.}



\begin{figure}[!tb] 
\centering 
\includegraphics[width=0.48\textwidth]{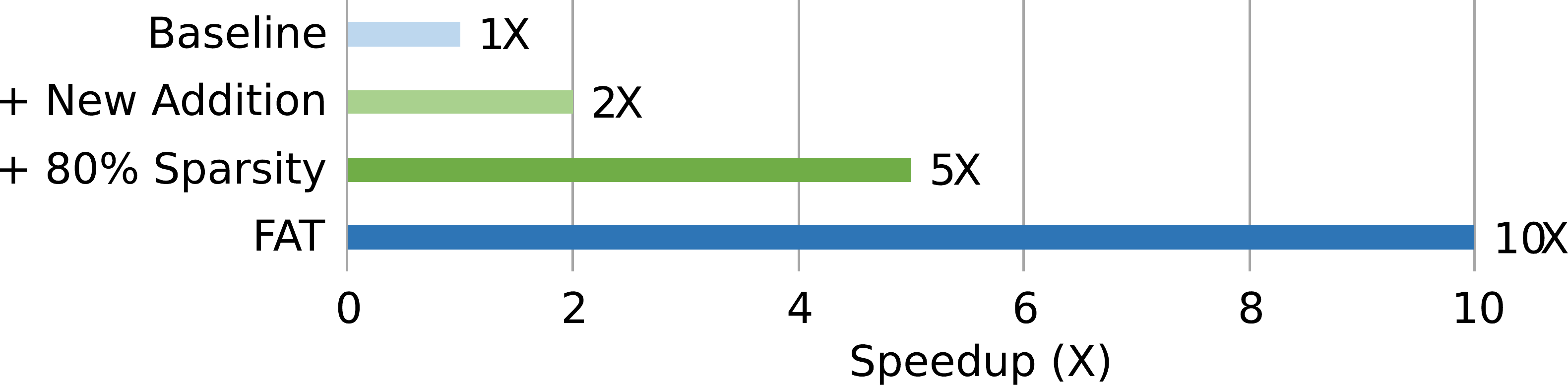} 
\caption{The speedup breakdown of TWNs with 80\% sparsity on our FAT.} 
\label{Fig-Speedup-Intro} 
\end{figure}


Our proposed TWN accelerator FAT combines the speedup from fast addition and sparsity together, as Fig.\ref{Fig-Speedup-Intro} shows. Taking the SOTA IMC BWN accelerator ParaPIM \cite{ASPDAC_2019_ParaPIM} as the baseline, the proposed fast addition scheme based on the Sense Amplifier provides \textcolor{black}{a} 2.00\(\times\) fast and 1.22\(\times\) power-efficient addition operator. Our Sparse Addition Control Unit provides another 5.00\(\times\) speedup with 80\% layer sparsity. \textcolor{black}{As a result,} FAT is up to 10.02\(\times\) faster and 12.19\(\times\) \textcolor{black}{more} energy-efficient than ParaPIM on quantized CNNs with 80\% \textcolor{black}{average sparsity}.

\section{Related Works and Motivation}

\subsection{Non-Volatile Memory and In-Memory-Computing}
Memory access is one of the \textcolor{black}{leading} performance and energy bottlenecks in traditional computing systems. Because moving the data from memory to processing units for computation and sending back the result to the memory lead to long latency and \textcolor{black}{massive} power consumption. Thus placing the logic near or even inside the memory have been proposed to mitigate this problem, namely Near-Memory Computing (NMC) \cite{DSD_2018_NMC, MM_2019_NMC} and In-Memory Computing (IMC) \cite{Nature_2020_IMC_devices,DATE_2018_CIM-Spin}. NMC places the logic closer to the memory, while IMC utilizes the memory arrays to do computation. 

IMC can be realized using Static Random-Access Memory (SRAM) \cite{TCS_2019_SRAM}, Dynamic RAM (DRAM) \cite{TCS_2019_DRAM}, and Non-Volatile Memories (NVMs) including Resistive RAM (ReRAM) \cite{ASP-DAC_2018_RRAM-BNN}, Phase Change Memory (PCM) \cite{Material_2021_PCM}, STT-MRAM \cite{DATE_2018_CIM-Spin}, Spin-Orbit Torque MRAM (SOT-MRAM) \cite{ASPDAC_2019_ParaPIM}, and other emerging memory devices \cite{Nature_2020_IMC_devices}. NVMs outperform SRAM and DRAM with near-zero leakage power and the non-volatile property of the stored data \cite{ICSICT_2018_SRAM-RRAM}. STT-MRAM and SOT-MRAM outperform ReRAM and PCM with \(\sim10^{15}\) memory cell write endurance and shorter write latency \cite{NVMW_2010_STT-MRAM-Endurance, ISVSI_2019_PiM}. \textcolor{black}{STT-MRAM also has excellent density, read speed and energy efficiency. ReRAM and PCM crossbars are widely adopted for matrix-vector multiplications, while STT/SOT-MRAM arrays perform parallel Boolean and addition operations. Therefore, this paper chooses STT-MRAM to accelerate addition-centric TWNs.}



\begin{figure}[!tb] 
\centering 
\includegraphics[width=0.48\textwidth]{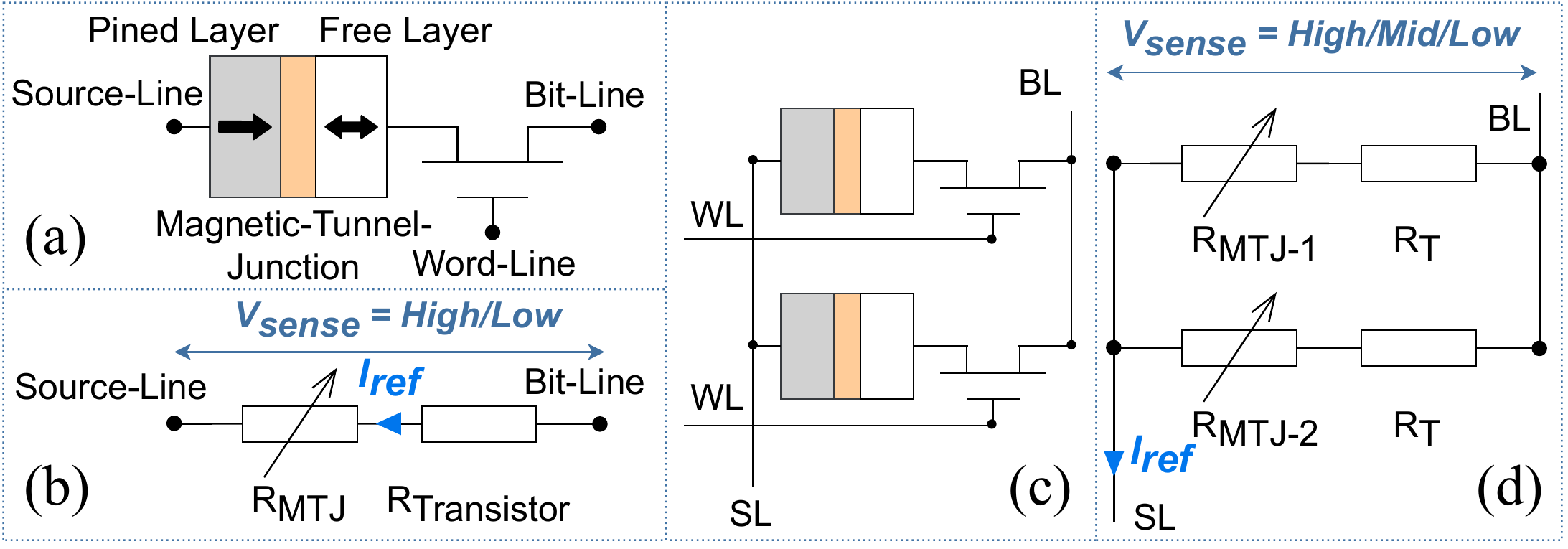} 
\caption{The STT-MRAM memory cells and the abstracted circuit. (a) Standard 1T-1MTJ STT-MRAM memory cell. (b) The equivalent circuit of one activated memory cell. (c) Activating two rows of memory cells simultaneously. (d) The equivalent circuit of two activated memory cells.} 
\label{Fig-stt-mram} 
\end{figure}

The standard STT-MRAM memory cell contains one Magnetic-Tunnel-Junction (MTJ) and one access transistor connected to the Bit-Line (BL), the Word-Line (WL), and the Source-Line (SL) as Fig.\ref{Fig-stt-mram} (a) shows. The MTJ has a pinned layer as the reference and a free layer whose magnetic orientation direction can be switched by memory writes. A tunnelling oxide isolates the pinned layer and the free layer. The MTJ has a low resistance in the "parallel state" where the free layer has the same magnetic orientation direction as the pinned layer, and it has a high resistance in the other "anti-parallel state". Applying the same current to the STT-MRAM memory cell, the Sense Amplifier receives a voltage \(V_{sense}\) as shown in Fig.\ref{Fig-stt-mram} (b). The anti-parallel state MTJ gives a high sensed voltage (can be identified as "1"), while the parallel state MTJ provides a low sensed voltage (identified as "0"). 
Similarly, In-Memory-Computing takes advantage of the resistance of the memory cells to do computation. For example, activating two rows of STT-MRAM memory cells simultaneously as Fig.\ref{Fig-stt-mram} (c) and (d) show, the sensed voltage can be high (the two cells storing "11"), middle (storing "01"/"10") or low (storing "00"). Thus the Sense Amplifier can perform computations like Boolean functions between the operands in the activated cells by identifying the sensed voltage, and this is the basic idea of In-Memory-Computing. 

\subsection{In-Memory-Computing Accelerators for Quantized CNNs} 


We review the related works on IMC accelerators for quantized neural networks in Table \ref{Tab-IMC-BNN-TNN}, \textcolor{black}{where traditional architecture based accelerators like Bit-Tactical \cite{ASPL_2019_TCL} are not listed.} As IMC architectures are able to compute Boolean functions and the addition operation, IMC based CNN accelerators mainly focus on ternary and binary networks: TNNs and BNNs, which quantize both the weights and activations to 2-bit/1-bit values and the main operations are Boolean operations, and TWNs and BWNs that only quantize the weights and the primary operation is addition. The accelerators can also be categorized into traditional memory based (SRAM and DRAM) and NVM based (including ReRAM, PCM, STT-MRAM, and SOT-MRAM) according to the memory type. 

BNNs replace the 1-bit multiplication with XNOR and the 1-bit accumulation with popcnt (count the number of "1"s in a binary value) in the binary dot product. \textcolor{black}{Thus} the BNN accelerators target XNOR and popcnt for high speedup. RRAM-BNN \cite{ASP-DAC_2018_RRAM-BNN}, XNOR-BNN \cite{ICSICT_2018_SRAM-RRAM} and PIMBALL \cite{TACO_2019_PIMBALL} utilize the XOR/XNOR operations of the ReRAM, SRAM and STT-MRAM memory arrays to process the BNNs. Some related works build special memory cells to accelerate BNNs. For example, 2T2R-TCAM \cite{ISCAS_2020_2T2R-TCAM-BNN} creates a 2-transistor-2-ReRAM (2T2R) Ternary Content Addressable Memory (TCAM) \textcolor{black}{that} supports in-memory logic and XNOR/XOR-based binary dot product. VR-XNOR \cite{JESTCAS_2019_VR-XNOR-BNN} proposes a memristor-based Voltage-Resistance XNOR (VR-XNOR) cell with a filter bank for binary convolution acceleration. Some papers also use traditional memories, especially SRAM, as processing elements for BNNs. SRAM-CIM \cite{TCS_2019_SRAM} \textcolor{black}{proposes} an SRAM Computing-In-Memory (SRAM-CIM) unit-macro for binarized fully connected neural networks. XNOR-SRAM \cite{JSSC_2020_XNOR-SRAM} presents an SRAM macro \textcolor{black}{supporting} ternary-XNOR-and-accumulate operations for both BNNs and TNNs. 
\begin{table}[tb]
\centering
\caption{Related IMC accelerators on binary and ternary networks.} 
\label{Tab-IMC-BNN-TNN} 
\begin{tabular}{|c|l|l|l|}
\hline
CNN                  & \multicolumn{1}{c|}{Target Ops}                                                                           & \multicolumn{1}{c|}{NVM Based} & \multicolumn{1}{c|}{Tradi. Mem. Based} \\ \hline
\multirow{5}{*}{BNN} & \multirow{5}{*}{\begin{tabular}[c]{@{}l@{}}XNOR +  Popcnt\end{tabular}}                                   & RRAM-BNN \cite{ASP-DAC_2018_RRAM-BNN}    & XNOR-SRAM \cite{JSSC_2020_XNOR-SRAM}     \\
                     &                                                                                                           & PIMBALL \cite{TACO_2019_PIMBALL}   & SRAM-CIM  \cite{TCS_2019_SRAM} \\
                     &                                                                                                           & XNOR-BNN \cite{ICSICT_2018_SRAM-RRAM}    & XNOR-BNN  \cite{ICSICT_2018_SRAM-RRAM}        \\
                     &                                                                                                           & 2T2R-TCAM \cite{ISCAS_2020_2T2R-TCAM-BNN}        &                                \\
                     &                                                                                                           & VR-XNOR \cite{JESTCAS_2019_VR-XNOR-BNN} &                                \\ \hline
\multirow{4}{*}{TNN} & \multirow{4}{*}{\begin{tabular}[c]{@{}l@{}}Ternary Mul., or \\ Gated-XNOR + \\ Popcnt\end{tabular}} & 4T2R-IM-DP \cite{JCAS_2021_4T2R-IM-DP}  & TiM-DNN \cite{TVLSI_2020_TiM-DNN}                   \\
                     &                                                                                                           & SpinLiM \cite{DATE_2021_SpinLiM}   &  XNOR-SRAM   \cite{JSSC_2020_XNOR-SRAM}           \\
                     &                                                                                                           & TeC-Cell \cite{DATE_2020_TeC-Cell}   & IMC-CD-TNN \cite{CICC_2021_IMC-CD-TNN}                               \\
                     &                                                                                                           & Ter-LiM  \cite{ISCAS_2021_Ter-LiM}         &                                \\  \hline
\multirow{2}{*}{BWN} & \multirow{2}{*}{\begin{tabular}[c]{@{}l@{}}Dense Addition\end{tabular}}                                   & ParaPIM \cite{ASPDAC_2019_ParaPIM}       &                                \\
                     &                                                                                                           & MRIMA \cite{TCAD_2019_MRIMA}            &                                \\ \hline
\multirow{2}{*}{TWN} & \multirow{2}{*}{\begin{tabular}[c]{@{}l@{}}Sparse Addition\end{tabular}}                               & \multirow{2}{*}{This Work}                 & \multirow{2}{*}{}              \\
                     &                                                                                                           &                                            &                                \\ \hline
\end{tabular}
\end{table}

TNN acceleration is another active research area besides BNN acceleration. The dominant operation in TNNs is ternary multiplication, which is equivalent to a Gated-XNOR followed by a popcnt. Thus most TNN accelerators build dedicated processing units for 2-bit ternary multiplications. 4T2R-IM-DP \cite{JCAS_2021_4T2R-IM-DP} provides a ReRAM-based IMC 4T2R bit-cell \textcolor{black}{to enable} In-Memory Dot Product (IM-DP) for TNNs. Ter-LiM  \cite{ISCAS_2021_Ter-LiM} implements a Ternary Logic-in-Memory (Ter-LiM) scheme based on the memristive dual-crossbar structure with multi-level memristor cells. SpinLiM \cite{DATE_2021_SpinLiM} utilizes two 2T-2MTJ SOT-MRAM cells to realize the ternary multiplication. TeC-Cell \cite{DATE_2020_TeC-Cell} proposes a non-volatile Ternary Compute-Enabled memory cell (TeC-Cell) based on ferroelectric transistors as well as TeC-Arrays to perform parallel signed ternary multiplications. The SRAM based TNN accelerators in related works also introduce special processing units for the ternary multiplications. For example, TiM-DNN \cite{TVLSI_2020_TiM-DNN} proposes Ternary-in-Memory (TiM) tiles for parallel signed ternary vector-matrix multiplications based on CMOS Ternary Processing Cells with memory and ternary multiplication functions. IMC-CD-TNN \cite{CICC_2021_IMC-CD-TNN} realizes the ternary multiplication with switch-capacitor ternary neurons and the wide vector summation in the charge domain to accelerate TNNs.

BWNs transform the multiplications into additions rather than the Boolean operations like BNNs and TNNs, so BWN accelerators utilize the addition and the memory array level parallelism in IMC \textcolor{black}{architectures}. ParaPIM \cite{ASPDAC_2019_ParaPIM} takes advantage of the parallel addition in SOT-MRAM memory arrays for efficient in-memory addition and convolution. \textcolor{black}{Similarly,} MRIMA \cite{TCAD_2019_MRIMA} leverages the massive parallelism across the STT-MRAM memory banks, matrices and arrays to accelerate BWNs. As a result, ParaPIM and MRIMA both achieve higher performance and energy efficiency than DRAM based, ReRAM based and ASIC based accelerators. 



\begin{figure*}[!tb] 
\centering 
\includegraphics[width=0.75\textwidth]{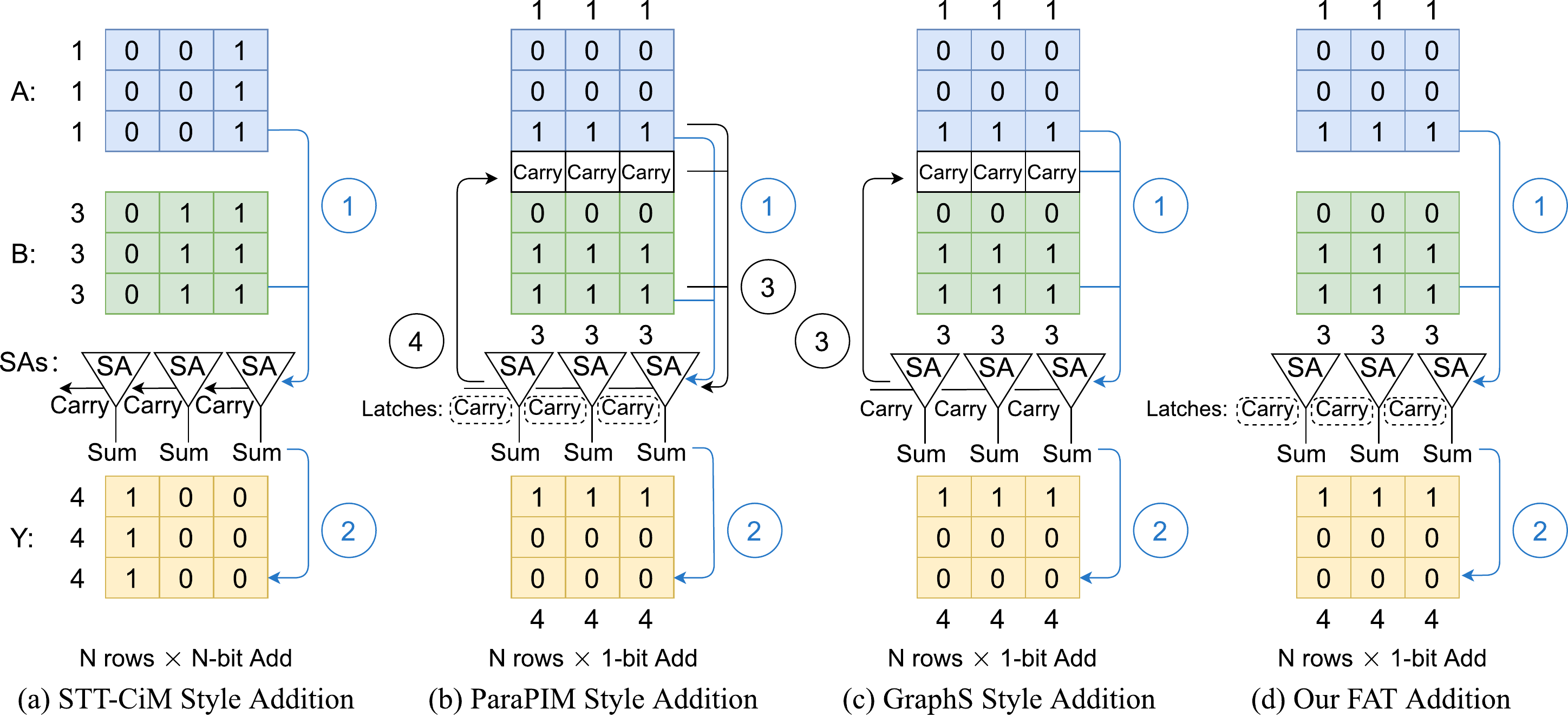} 
\caption{\textcolor{black}{Addition schemes in STT-CiM series \cite{TVLSI_2017_STT-CiM, ISLPED_2019_FEFET-CiM, TN_2020_VCS-CiM}, ParaPIM series \cite{ASPDAC_2019_ParaPIM, TCAD_2019_MRIMA},  GraphS series \cite{DATE_2019_GraphS, NanoArc_2019_ParaPIM,ASP-DAC_2020_CA-DNN-PIM} IMC designs and our proposed FAT.}}
\label{Fig-add} 
\end{figure*}

TWNs are different from BWNs due to the 2-bit weights, the higher accuracy and the network sparsity. To the best of our knowledge, this work is the first sparse IMC accelerator for TWNs. This work builds more efficient addition operations than ParaPIM and MRIMA. This paper also takes the 2-bit representation and the network sparsity into account to enable sparse additions for lower latency and higher energy efficiency.

\subsection{In-Memory-Computing Addition Schemes}
Existing IMC devices are able to perform Boolean functions and Addition operations. The mainstream IMC addition schemes can be categorized into STT-CiM series, ParaPIM series and GraphS series, as Fig. \ref{Fig-add} shows. We will review and analyze the addition schemes in these related works and propose our efficient addition scheme in this subsection.

STT-CiM series \textcolor{black}{designs} \cite{TVLSI_2017_STT-CiM, ISLPED_2019_FEFET-CiM, TN_2020_VCS-CiM} store the values in rows, as Fig. \ref{Fig-add} (a) shows. The upper two arrays stand for the operands A=(1, 1, 1) and B=(3, 3, 3). First, the Sense Amplifier (SA) of STT-CiM reads two operands together and analyzes the total current to calculate the summation result Y=A+B. Second, the SA outputs the addition result Y=(4, 4, 4), and the result can be stored in the memory array at the bottom. This addition scheme is logically straightforward and efficient for scalar addition. However, it has two shortcomings: First, it needs to wait for the carry to propagate to the last bit, \color{black}so the scalar latency \(ts_{STT}\) is long. As equation (1) shows, \(ts_{STT}\) increases linearly with the operand bitwidth N due to the carry calculation and propagation latency \(t_{Carry}\). Second, the vector addition \(tv_{STT}\) has N times latency as the scalar addition, as equation (2) shows.
\begin{align}
    ts_{STT} & \color{black} = t_{Read} + (N-1) \cdot t_{Carry} + t_{SUM} + t_{Write}  \\
    \color{black}
    tv_{STT} & \color{black} = ts_{STT} \times N \\
    \color{black}
    tv_{FAT} & \color{black} = (t_{Read} + t_{SUM}  + t_{Write}) \times N
\end{align}


    


\color{black}

ParaPIM \cite{ASPDAC_2019_ParaPIM} and MRIMA \cite{TCAD_2019_MRIMA} are IMC accelerators for BWNs based on SOT-MRAM and STT-MRAM, respectively. ParaPIM series accelerators store the values in columns, as Fig. \ref{Fig-add} (b) shows. For example, operand 1 encoded as "001" is stored in a column rather than a row in the first memory array A. ParaPIM series accelerators perform the addition bit by bit rather than finish the scalar addition in one step. \textcolor{black}{This addition scheme has the advantage that the vector addition has the same latency as the scalar addition as long as the vector length does not exceed the number of memory columns. However,} it also has two weaknesses: First, it computes the Sum and Carry-out sequentially, which brings long latency. Second, this addition scheme needs to write the Carry-out back to the memory to utilize it as the Carry-in for the next bit, which results in extra latency and energy cost.

GraphS \cite{DATE_2019_GraphS}, ParaPIM-SA-II \cite{NanoArc_2019_ParaPIM} and CA-DNN-PIM \cite{ASP-DAC_2020_CA-DNN-PIM} adopt a new SA design to overcome the first weakness of ParaPIM series addition while keeping a similar workflow. GraphS series \textcolor{black}{designs} compute the Sum and Carry-out in one step as Fig. \ref{Fig-add} (c) shows and achieve 1.3\(\times\) performance/area as ParaPIM. \textcolor{black}{However,} GraphS series \textcolor{black}{designs} still have two problems: First, GraphS only has 0.8\(\times\) energy-efficiency/area as ParaPIM due to its complex 3-operand logic and the increased area. Second, GraphS series accelerators still store back and read in the intermediate carry results to/from the memory array following ParaPIM, bringing unnecessary latency and energy.

To overcome the shortcomings of existing addition schemes, we propose an efficient addition scheme at the algorithm level along with a novel SA at the circuit level. Our FAT style fast addition operation does not need to wait for the propagation of the carry nor write the carry back to the memory, as illustrated in Fig. \ref{Fig-add} (d). First, similar to ParaPIM and GraphS, we store the operands in columns and compute the summation bit by bit. Second, instead of storing the intermediate carry of the previous bit back to the memory array, we keep it into a D-latch inside the SA to make it more energy-efficient. Third, we compute the summation by reading in two operands in only one step with the Carry-in (same as the Carry-out from the previous bit) stored inside the SA, which brings a shorter critical path for addition operations. \textcolor{black}{As the carry is used in the next bit, storing the carry in a latch hides the carry calculation and propagation latency. As equation (3) shows,  FAT's vector addition latency \(tv_{FAT}\) equals N times 1-bit addition, shorter than STT-CiM when N\textgreater 1.} 
Also, we only need 2-operand logic instead of the 3-operand logic in ParaPIM and GraphS, reducing the power and increasing the sensing reliability.


\section{Accelerator FAT}
We present our accelerator FAT in this section, including the architecture overview, \textcolor{black}{the memory arrays with computational capacity,} and the data mapping method. The memory array features two components: the Sparse Addition Control Unit (SACU), which provides a sparse vector dot product, and the Sense Amplifier (SA), which implements the proposed fast addition scheme. In addition, the Combined-Stationary data mapping method reduces the data movement and increases the parallelism of memory arrays.




\subsection{Overview}
Our accelerator, FAT, is an STT-MRAM based In-Memory-Computing (IMC) accelerator for Ternary Weight Neural Network (TWN) inference. FAT also supports Binary Weight Neural Network (BWN) inference with few configurations. We will explain the accelerated networks, the accelerator architecture and the basic workflow in this subsection.

\subsubsection{Target Applications}
The typical basic block of \textcolor{black}{CNNs} comprises a convolutional or fully connected layer followed by an activation function and a batch normalization layer. The convolutional and fully connected layers are computation-intensive, \textcolor{black}{while the activation and batch normalization layers only have a few floating-point operations}. For example, \textcolor{black}{the} convolution of activation \(X\) and weight \(W\) needs \(C\cdot KH\cdot KW\) multiplications for every output point, as equation (\ref{eq-conv}) shows, where \(KN,\ C,\ KH,\ KW\) are the number of filters, channel, kernel height, and kernel width. \textcolor{black}{In comparison, the popular activation function ReLU() and the batch normalization BN() are much simpler than convolution}, as equations (\ref{eq-relu})-(\ref{eq-bn}) show. 
\begin{align}
\label{eq-conv}
Y_{kn, h, w} &=\sum_{c=1}^{C}\sum_{i=1}^{KH}\sum_{j=1}^{KW}X_{c,h+i,w+j} \cdot W_{kn, c,i,j}  \\
\label{eq-relu}
ReLU(x) &=\left\{\begin{matrix}
x, & x>0\\ 
0, & x \leq 0
\end{matrix}\right. \\
\label{eq-bn}
BN(Y) &=\frac{Y-\textup{E}[Y]}{\sqrt{\textup{Var}[Y]+\epsilon}}
\end{align}


The weights in TWNs are quantized to ternary values to reduce the computation complexity of convolution and fully connected layers. The weights are ternarized to \{+1, 0, -1\} by comparing with trained or given thresholds, as equation (\ref{eq-w-tern}) shows, where \(w\) and \(w^t\) are the original and ternarized weight values, and \(TH_{low}\) and \(TH_{high}\) are the thresholds. Modern TWNs train the weights to be ternary values \cite{ICLR_2017_TTQ,AAAI_2020_RTN,AAAI_2021_TRQ}, \textcolor{black}{thus} the weights of target neural networks are already quantized into 2-bit numbers when the training finishes. 
\begin{align}
\label{eq-w-tern}
w^{t} &= \left\{\begin{matrix}
+1, w > TH_{high} &\ \\ 
-1, w < TH_{low}\ &, TH_{low} < TH_{high}\\ 
\ 0, otherwise & \ 
\end{matrix}\right. \\
\label{eq-xt-wt}
y &=\vec{x} \cdot \vec{w}^t = \sum_{i=1}^{N} \vec{x}_i \cdot \vec{w}_i^t \ , \ \vec{w}^t\in \{\pm 1,0\}^N 
\end{align}

The convolution and matrix multiplication in convolution and fully connected layers can be decomposed as the inner dot product between the activation vector \(\vec{x}\) and the weight vector \(\vec{w}^t\), as equation (\ref{eq-xt-wt}) shows, where \(N\) is the vector length. \textcolor{black}{Furthermore,} the ternarized weights in TWNs simplify the dot product into addition/subtraction and null operations (multiplying zero is non-sense). Therefore, we build FAT to accelerate TWNs with two main features: the low-level fast addition and the high-level addition based sparse dot product \textcolor{black}{that} skips the null operations. FAT is also able to \textcolor{black}{infer} BWNs because the weights of BWNs only contain \{+1, -1\}, but BWN inference has no performance benefit from the sparsity.

\subsubsection{Accelerator Architecture}

\begin{figure}[!b] 
\centering 
\includegraphics[width=0.48\textwidth]{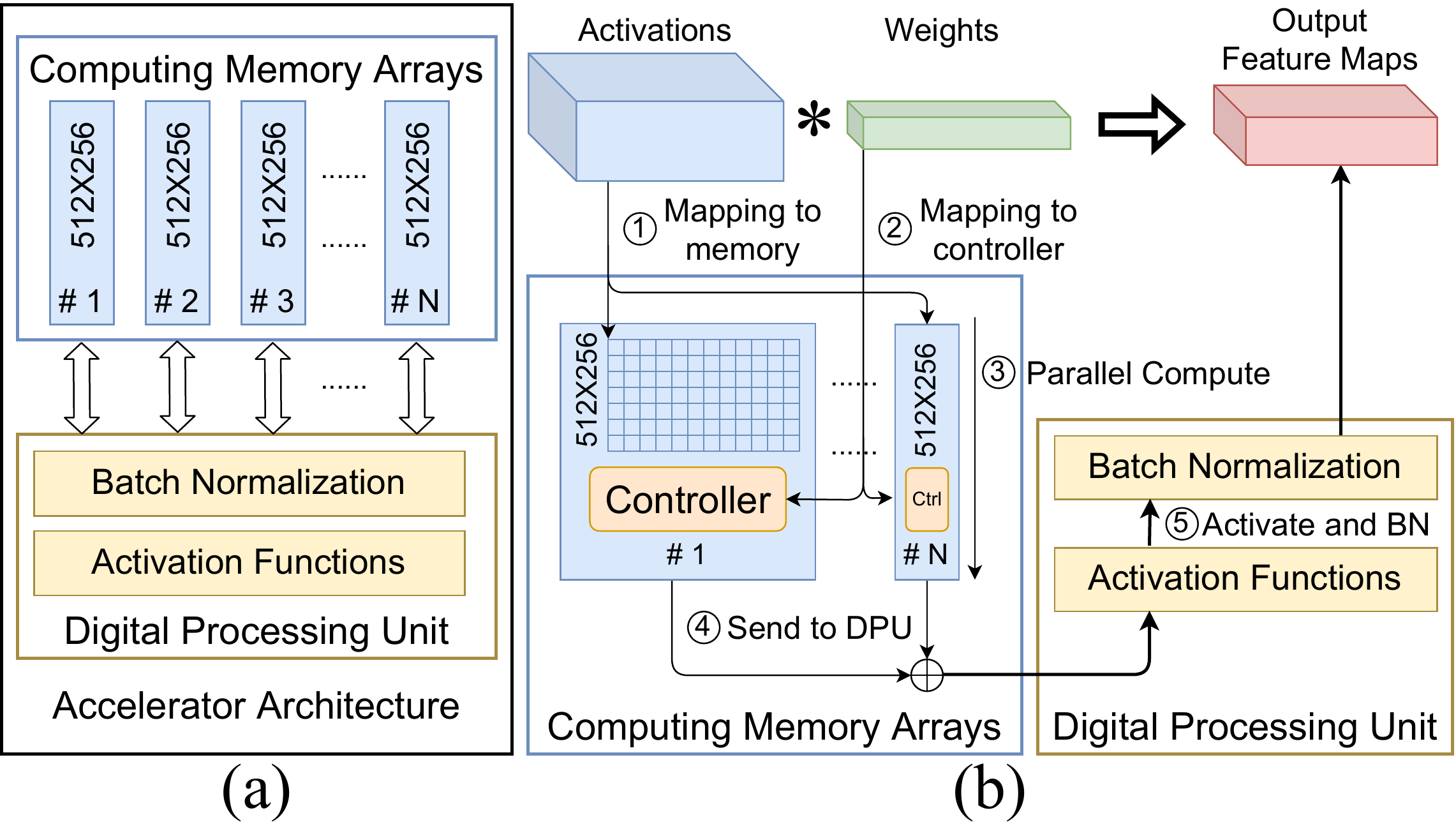} 
\caption{(a) The architecture of FAT. (b) The computation workflow of FAT.} 
\label{Fig-workflow} 
\end{figure}

\begin{figure*}[!tb] 
\centering 
\includegraphics[width=0.98\textwidth]{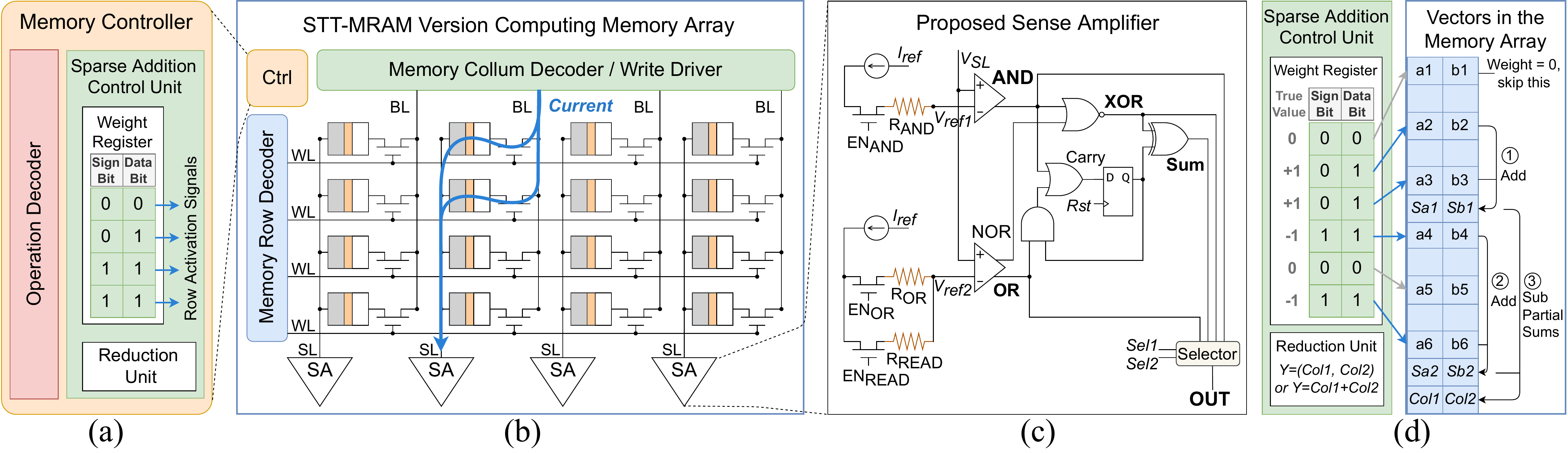} 
\caption{Overview of FAT Computing Memory Array. (a) The architecture of the Memory Controller. (b) The architecture of the Computing Memory Array. (c) The architecture of the Sense Amplifier. (d) The workflow of addition based sparse dot product controlled by the Sparse Addition Control Unit.} 
\label{Fig-arch} 
\end{figure*}

Our accelerator FAT consists of Computing Memory Arrays (CMAs) and a CMOS Data Processing Unit (DPU) as Fig. \ref{Fig-workflow} (a) shows. \textcolor{black}{The CMAs are in charge of the addition operations in convolutional and fully connected layers, equipped with proposed SACU and SA for sparse dot products and efficient addition operations. FAT contains 4096 CMAs with 64MiB total memory capacity and 128K 2-bit weight registers.} 

The DPU takes care of the batch normalization and activation layers. We keep almost the same DPU architecture as ParaPIM \cite{ASPDAC_2019_ParaPIM} and MRIMA \cite{TCAD_2019_MRIMA} except that our DPU has no hardware quantizer. As the weights of modern TWNs and BWNs are trained to be 2-bit ternary values or 1-bit binary values, which means the weights are already quantized before inference, \textcolor{black}{the weight quantizer is unnecessary.} Excluding the quantizer \textcolor{black}{also reduces} the chip area, power, and inference time thanks to no procedure on quantization. In summary, the CMAs with sparse dot product and efficient addition work together with the DPU containing activation functions and batch normalization to finish the TWN inference.




\subsubsection{Computation Workflow}


FAT conducts the inference block by block, and the workflow of one example convolution block is shown in Fig. \ref{Fig-workflow} (b). \underline{First}, the activations of each convolution layer are loaded to the memory cells of CMAs. \underline{Second}, the weights of each layer are loaded to the Sparse Addition Control Unit inside the Memory Controller. The loading of weights may repeat many times as there are many filters in a convolution layer. \underline{Then} the Memory Controller performs parallel computation across memory columns. All the CMAs with loaded activations and weights can process the convolution layer in parallel. \underline{Fourth}, the convolution results are sent to the DPU through the internal buses. \underline{Fifth}, the DPU performs the activation function and batch normalization to generate the output feature maps. \textcolor{black}{After all, the output feature maps are stored again from the DPU to CMAs to work as activations of the next layer. We repeat this process until we get the output of the whole neural network.} 


\subsection{Computing Memory Array}
The Computing Memory Array (CMA) consists of a Memory Controller (MC), a Memory Row Address Decoder (MRAD) which controls the Word-Lines (WLs), a Memory Column Address Decoder (MCAD)/Writer Driver that controls the Bit-Lines (BLs), a memory array, and Sense Amplifiers (SAs) connected to the Source-Lines (SLs), as Fig. \ref{Fig-arch} shows. \textcolor{black}{Keeping the same array size as related works \cite{ASPDAC_2019_ParaPIM,ASP-DAC_2020_CA-DNN-PIM}, each CMA has 512 rows and 256 columns. FAT adopts the same column-major data storage format as ParaPIM series \cite{ASPDAC_2019_ParaPIM, TCAD_2019_MRIMA} and GraphS series \cite{DATE_2019_GraphS, NanoArc_2019_ParaPIM,ASP-DAC_2020_CA-DNN-PIM} \textcolor{black}{designs}, which stores different bits of one number in sequential rows in one column. As we store all the data in column-major format, there is no need to convert the data between bit-serial and bit-parallel.} 

Our CMA works in three modes: a standard memory device mode with basic Read/Write support, \textcolor{black}{a traditional IMC device mode that enables the Boolean and Addition functions}, and a TWN accelerator mode. \textcolor{black}{Working as a standard memory device or a traditional IMC device, the MC receives instructions from a CPU and sends control signals to other components to perform Read, Write, Boolean and Addition operations.} The MRAD/MCAD decodes the row/column addresses based on the signals from MC and activates the WLs/BLs to access target memory cells. Then the SA receives the current from the memory array by the SL and produces the corresponding result of Read, Boolean functions or Addition. \textcolor{black}{These two memory modes also utilize the non-volatility of STT-MRAM by keeping the data after power-off and avoiding data reloading when powered on again.} Working as a TWN accelerator, our CMA provides two main features. First, the Sparse Addition Control Unit (SACU) inside the MC enables the high-level sparse vector dot product. Second, the SA facilitates the low-level fast addition. We will introduce the SACU and the SA in the following two subsections in detail. 




\subsubsection{Sparse Addition Control Unit}
When processing a convolution layer or a fully connected layer in TWNs, the weights determine which operands of the activations should be added together. Therefore, we store the activations inside the memory array to conduct Addition operations. At the same time, we load the weights into the MC, to be exact, the inside Sparse Addition Control Unit (SACU). 

\paragraph{Architecture}
The architecture of the SACU is shown in Fig. \ref{Fig-arch} (a). It contains weight registers and a reduction unit. The SACU uses weight registers to generate the control signals of the sparse dot product. The reduction unit can accumulate the summations in different columns when necessary.

The weights of TWNs only have three values, namely \{+1, 0, -1\}. Thus we adopt the encoding of standard signed integer for the weights, as Table \ref{Tab-weights} shows. The 2-bit encoding \textcolor{black}{containing} a sign bit and a data bit \textcolor{black}{determines what operation} should be applied to the activations in corresponding rows. The sign bit shows whether this is an addition (sign bit=0) or a subtraction (sign bit=1). As +1(01) and 0(00) share the same sign bit "0", we need the data bit to work as a mask. Therefore, we activate the corresponding rows only when data bit=1 so that the rows corresponding to weight 0(00) have no operation. In other words, the null operations are skipped during the sparse dot product.
\begin{table}[]
\centering
\caption{Utilizing the weights to skip the null operations.} 
\label{Tab-weights}
\begin{tabular}{| r | c  c | c  c |}
\hline
\multicolumn{1}{|c|}{Weight} & Sign Bit & Add/Sub & Data Bit/Mask & Activate this row? \\
\hline
+1 (01)                    & 0        & Add     & 1               & Yes                \\
\hline
0 (00)                     & 0        & Null    & 0               & No                 \\
\hline
-1 (11)                    & 1        & Sub     & 1               & Yes              \\
\hline
\end{tabular}
\end{table}

\paragraph{Sparse Addition Workflow}
As mentioned in the previous paragraph, the data bit of the weight determines whether the operands in the corresponding row need operation and the sign bit indicates the operation type is addition or subtraction. This motivates us \textcolor{black}{to conduct the addition-based sparse dot product following Fig. \ref{Fig-arch} (d) workflow.} The SACU generates the row activation signals based on the data in \textcolor{black}{weight registers, and} the MRAD enables the corresponding rows of operands for addition. 

\textcolor{black}{The addition based sparse dot product has three stages. Taking Fig. \ref{Fig-arch} (d) as an example, two vectors \(a\) and \(b\) in the memory array perform dot product with the weight vector (0, +1, +1, -1, 0, -1). \underline{First}, we add the operands corresponding to weight +1 and get a partial sum in each column. The SACU activates row 2 and row 3 simultaneously, adds them together, and produces partial sums \(Sa1\) and \(Sb1\). \underline{Second}, we add the operands corresponding to weight -1 and get another partial sum in every column. The SACU adds row 4 and row 6 together and calculates partial sums \(Sa2\) and \(Sb2\) simultaneously. \underline{Third}, a subtraction operation (SUB=NOT+ADD) between these two partial sums produces the summation result of the whole column. The SACU performs \(Col1=Sa1-Sa2\) and \(Col2=Sb1-Sb2\) in parallel.} 

\textcolor{black}{Our sparse dot product has three features. First, it automatically skips the addition operations corresponding to zero weights. Because the SACU only activates the rows where the weights are +1 or -1. Second, it has outstanding memory column level parallelism. As the example shows above, the memory columns can be activated simultaneously to conduct parallel processing across the whole memory array. We will combine this column level parallelism with suitable mapping and scheduling to boost the performance further in the Data Mapping section. Third, it replaces the subtractions corresponding to weights -1 with additions to achieve higher performance and energy efficiency. The addition is more efficient than subtraction on FAT because subtraction is realized by NOT and addition. The proposed three-stage addition pipeline separates the operands corresponding to weight +1 and -1, and performs additions on all the activation operands and only one subtraction on the partial sums, which reduces the dot product latency and energy.}

\textcolor{black}{As mentioned in the accelerator overview,} FAT could also work as a BWN accelerator with simple configurations. We only need to extend the 1-bit binary weight data (+1 and -1) to 2-bit signed integers \textcolor{black}{according to Table \ref{Tab-weights}}, then store the 2-bit binary values into weight registers in the SACU, and FAT will work as a BWN accelerator. However, the SACU will activate all the rows based on the weights in this case, so there will be no benefit from sparsity.  

\subsubsection{Sense Amplifier}
The Sense Amplifier (SA) is the core component of an In-Memory-Computing (IMC) device because it determines which kinds of operations the device support. Our motivation drives us to design an SA that supports faster and more efficient addition operations than related works. Furthermore, our SA is able to conduct common Boolean functions and subtraction not limited to the addition operation. \textcolor{black}{In the following paragraphs,} we will introduce the SA architecture, the supported operations, and the comparison with related works.

\paragraph{Architecture and Workflow}
Our SA's architecture is presented in Fig. \ref{Fig-arch} (c). The signal flow inside the SA has four stages: sensing, comparing, combining, and selecting. We will follow the signal flow to show how our SA works. 

\begin{figure}[!tb] 
\centering 
\includegraphics[width=0.48\textwidth]{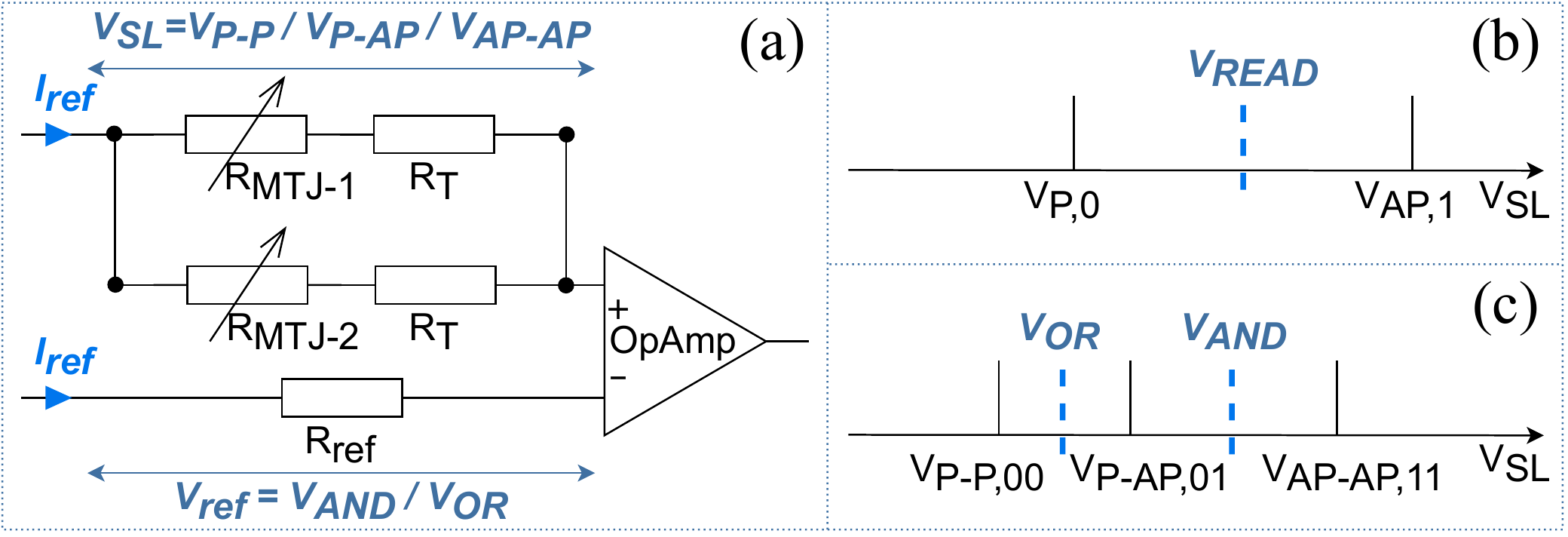} 
\caption{(a) The equivalent circuit when sensing two memory cells at the same time. (b) The reference voltage \(V_{READ}\) of sensing one memory cell. (c) The reference voltages \(V_{AND}\) and \(V_{OR}\) of sensing two memory cells.} 
\label{Fig-SA-voltage} 
\end{figure}

\textbf{Sensing}: In the sensing stage, the Word-Line (WL) and the Bit-Line (BL) are activated, and the current flows through the target memory cells to the Source-Line (SL), as Fig. \ref{Fig-arch} (b) shows. The equivalent circuit of the sensing state is illustrated in Fig. \ref{Fig-SA-voltage} (a). The Operational Amplifier (OpAmp) in the SA gets the total current from the SL along with the voltage of SL \(V_{SL}\), as equation (9) shows, where \(I_{ref}\) is the reference activation current, and \(R_{MTJ}\) and \(R_{T}\) are the resistances of the MTJ and the access transistor in a memory cell.
\begin{align}
  V_{SL} &= I_{ref}\cdot((R_{MTJ1}+R_{T1})//(R_{MTJ2}+R_{T2})) \\
  V_{ref}&= I_{ref}\cdot R_{ref}
\end{align}




\textbf{Comparing}: Getting the sensed voltage \(V_{SL}\), the OpAmp in the SA compares it with reference voltages \(V_{ref}\) (generated in equation (10)) to get the result of READ, AND or OR, which can be further used to compute more complex functions. The sensed voltage \(V_{SL}\) of reading out a single memory cell can be either a lower voltage \(V_{P,0}\) when the MTJ is in \textcolor{black}{a} parallel state (storing a "0") or a higher voltage \(V_{AP,1}\) with an anti-parallel state MTJ (storing a "1"). \textcolor{black}{Thus} the reference voltage for reading \(V_{READ}\) lies between \(V_{P,0}\) and \(V_{AP,1}\), as Fig. \ref{Fig-SA-voltage} (b) shows. Similarly, the sensed voltage \(V_{SL}\) of reading out two memory cells can be \(V_{P-P,00}\), \(V_{P-AP,01}\) and \(V_{AP-AP,11}\), as Fig. \ref{Fig-SA-voltage} (c) shows. \textcolor{black}{Thus} the reference voltage of AND \(V_{AND}\) lies between \(V_{P-AP,01}\) and \(V_{AP-AP,11}\), while the \(V_{OR}\) is between \(V_{P-P,00}\) and \(V_{P-AP,01}\). 
\begin{align} 
    A\ \textup{XOR}\ B&=[\underline{A\ \textup{AND}\ B}]\ \textup{NOR}\ [\underline{A\ \textup{NOR}\  B}]  \\
    SUM &= [\underline{A\ \textup{XOR}\ B}]\ \textup{XOR}\ Cin \\
    Cout &= ([\underline{A\ \textup{OR}\ B}]\ \textup{AND}\ Cin)\ \textup{OR}\ [\underline{A\ \textup{AND}\ B}]
\end{align}

\textbf{Combining}: Next, the other logic gates in the SA combine the AND, OR and NOR signals generated by the OpAmps to compute more complex functions, e.g., XOR, SUM and Carry-out of the addition operation as presented in Fig. \ref{Fig-arch} (c). For example, the XOR is calculated by NOR between the AND and NOR of operands A and B, as equation (11) shows, where the underlined parts in the square brackets are signals generated by the OpAmp from the comparing stage. Similarly, the SUM and the Carry-out \(Cout\) are calculated following equation (12)-(13) where \(Cin\) is the Carry-in from the previous bit stored in the D-Latch. Thus our SA uses four logic gates (NOR, XOR, OR and AND) and one D-Latch in the combining stage, as equation (11)-(13) and Fig. \ref{Fig-arch} (c) show.

\textbf{Selecting}: Finally, the selector selects the desired result based on the selecting signals Sel1 and Sel2, and sends it to the output port OUT. 


\begin{table}[!tb]
\centering
\caption{Configuration of enable signals of the Sense Amplifier.} 
\label{Tab-EN-signals}
\setlength\tabcolsep{5pt} 
\begin{tabular}{|l|c|c|c|c|c|c|c|}
\hline
Operation   & READ & NOT & AND & NAND & OR & XOR & ADD \\ \hline
EN\_READ      & 1    & 0   & 0   & 0    & 0  & 0   & 0   \\ \hline
EN\_AND     & 0    & 1   & 1   & 1    & 0  & 1   & 1   \\ \hline
EN\_OR      & 0    & 1   & 0   & 0    & 1  & 1   & 1   \\ \hline
Selector Port & OR   & XOR & AND & XOR  & OR & XOR & SUM \\ \hline
\end{tabular}
\end{table}

\begin{table}[!b]
\centering
\caption{Configuration of selector signals of the Sense Amplifier.} 
\label{Tab-Sel-signals}
\begin{tabular}{|c|c|c|c|c|}
\hline
Selector Port & AND & OR & XOR & SUM \\ \hline
Sel 1         & 0   & 0  & 1   & 1   \\ \hline
Sel 2         & 0   & 1  & 0   & 1   \\ \hline
\end{tabular}
\end{table}

\paragraph{Configuration and Supported Operations}
Our SA performs READ, NOT, AND, NAND, OR, XOR, and Addition (ADD) functions natively and Subtraction (SUB) extensively. It relies on the enable signals and selector signals from the Memory Controller (MC) to give the desired result. We configure the enable signals of the SA to perform different functions according to Table \ref{Tab-EN-signals}. Meanwhile, we select the results routed to the input ports of the selector as Table \ref{Tab-Sel-signals} shows, \textcolor{black}{and} then we get the desired result at the OUT port.

The SA supports eight functions but building eight selector ports for all the functions brings high complexity and large area cost. \textcolor{black}{Thus, we optimize the SA on READ, NOT, NAND and SUB functionality to simplify the design to only four selector ports.} \underline{First}, the READ and OR operations share the same OR selector port because they use the same OpAmp. \underline{Second}, as the NOT equals to XOR with "1"s as equation (14) shows, we read in the operand along with a row filled with "1"s and produce the NOT result at the XOR selector port. RISC-V has adopted the same technique to simplify the instruction set by replacing NOT with XOR \cite{RISC-V}. \underline{Third}, we disable the enable signals of EN\_OR and EN\_READ at the second OpAmp in the SA when computing the NAND. \textcolor{black}{Then the OpAmp NOR port produces "0"s for any \(V_{SL}\) higher than zero.} The NAND result appears at the XOR port after a NOR between the AND and "0"s, as equation (15) shows. \underline{Last}, as the SUB equals to ADD the opposite number as shown in equation (16), we perform the SUB operation by one NOT followed by one ADD with the first carry-in to be "1".
\begin{align}
    \textup{NOT}\ A &=A\ \textup{XOR}\ 111...1 \\
    A\ \textup{NAND}\ B&=(A\ \textup{AND}\ B)\ \textup{NOR}\ 000...0  \\
    A - B &= A + ((\textup{NOT}\ B) + 1) 
\end{align}

Therefore, our SA has the least number of enabling (EN) signals, selector signals and amplifiers among related works, as inferred from Table \ref{Tab-SA-compare}. \textcolor{black}{Nevertheless, the XOR and addition operations cost one D-latch and four Boolean gates.} The area of the SAs can also be inferred from Table \ref{Tab-SA-compare}. \textcolor{black}{STT-CiM's SA has} smaller area than ours due to one less D-latch, but STT-CiM has four more control signals than FAT. The SA of FAT \textcolor{black}{has} a smaller area than ParaPIM's SA thanks to fewer control signals and a smaller output selector. Operational amplifiers usually have a much larger area than Boolean gates due to higher complexity and higher loading capacity. \textcolor{black}{Thus} the SA of FAT will be smaller than GraphS' SA, which has one more operational amplifier and a larger selector. The exact area and speed comparison will be given in the Evaluation section.

\begin{table}[tb]
\centering
\caption{Comparison of proposed SA and SAs in related works.} 
\label{Tab-SA-compare} 
\begin{tabular}{|c|c|c|c|c|c|}
\hline
\multirow{2}{*}{\textcolor{black}{Designs}} & \multicolumn{2}{c|}{Signals}    & \multicolumn{3}{c|}{Circuits} \\ \cline{2-6} 
                      & EN           & Sel.         & Amplifier  & D-Latch  & Boolean Gates  \\ \hline
STT-CiM \cite{TVLSI_2017_STT-CiM}              & 6            & 3                & 2          & 0        & 4                   \\ \hline
ParaPIM  \cite{ASPDAC_2019_ParaPIM}             & 4            & 3                & 2          & 1        & 3                   \\ \hline
GraphS   \cite{DATE_2019_GraphS}             & 6            & 3                & 3          & 0        & 1                   \\ \hline
\textcolor{black}{\textbf{Our FAT}}               & \textcolor{black}{\textbf{3}}            & \textcolor{black}{\textbf{2}}                & \textcolor{black}{\textbf{2}}          & \textcolor{black}{\textbf{1}}        & \textcolor{black}{\textbf{4}}                   \\ \hline
\end{tabular}
\end{table}




\paragraph{Fast Addition}
We propose the new SA to adopt the efficient addition scheme shown in Fig. \ref{Fig-add} (d). Temporarily our CMA works as a sequential 1-bit adder and computes the summation bit by bit to realize the N-bit addition. \underline{First}, the MC initializes the D-Latch containing the carry in the SA as 0 before the addition operation. \underline{Next}, the MCAD enables those columns containing the operands, and the MRAD activates two rows catering to two bits of the operands simultaneously. The reference current flows through the memory cells to the SA. \underline{Then} the SA computes the summation (SUM) of these two bits and stores the carry-out of the current bit in the Carry Latch to use it as the carry-in of the next bit. This unique design avoids storing back the carry-out results to the memory and saves a considerable amount of \textcolor{black}{time and} energy compared with ParaPIM \cite{ASPDAC_2019_ParaPIM} and GraphS \cite{DATE_2019_GraphS} series IMC accelerators. \textcolor{black}{Thanks to the bit-by-bit addition, storing the carry in a latch also hides the latency of carry calculation and propagation in the STT-CiM \cite{TVLSI_2017_STT-CiM} series designs.} 
\textcolor{black}{Compute-SRAM \cite{JSSC_2019_Compute-SRAM} is an SRAM based IMC \textcolor{black}{design} that adopts similar one-step bit-serial addition and storing the carry inside a D-Latch as FAT. Except for memory specific differences, FAT distinguishes from Compute-SRAM on using OR instead of XOR to generate the carry and simplifying the output selection to use a 4-input selector rather than an 8-input one. FAT has faster carry signal than Compute-SRAM because OR is earlier than XOR in the SA as equations (11)-(13) show. FAT has faster SUM signal than Compute-SRAM because our XOR port has fewer loading circuits than Compute-SRAM's and the 4-input selector has shorter latency than the 8-input selector.} 

\textcolor{black}{Our addition operation is throughput driven rather than latency driven. FAT has longer latency than STT-CiM series IMC designs on single scalar addition} because we compute the summation bit by bit. FAT's advantage lies in the vector addition, where hundreds to thousands of pairs of operands need addition results. Thanks to the column-major data storage format shown in Fig. \ref{Fig-add}, the bit-by-bit addition in FAT, ParaPIM and GraphS has N times parallelism as STT-CiM series \textcolor{black}{designs} on N-bit vector addition. \textcolor{black}{Supposing two 1024\(\times\)32-bit vectors stored in 1024 columns and 32 rows conduct addition inside a memory array.} FAT repeats the 1-bit addition 32 times to get the vector addition result because the operand vector is stored in 1024 columns vertically. \textcolor{black}{In contrast,} STT-CiM performs the 32-bit addition 32 times across the memory array because the operand vector is stored in 32 rows horizontally. \textcolor{black}{Our FAT has shorter vector addition latency than STT-CiM because FAT conducts the 1-bit addition 32 times while STT-CiM repeats the 32-bit addition 32 times.} 

\textcolor{black}{In summary, our addition scheme is faster than ParaPIM, GraphS, and Compute-SRAM on scalar and vector addition, and is faster than STT-CiM on vector addition.} Thus we refer to the proposed addition scheme as fast addition. The Evaluation section will provide a detailed performance comparison of these works. As the IMC addition operands are integers or fixed-point numbers, the TWN activations also need to be stored as integers rather than floating-point numbers. \textcolor{black}{Other types of memories may adopt the proposed fast addition scheme, given that they can read two memory rows simultaneously and distinguish "00", "01"/"10" and "11" inside sense amplifiers. Therefore, adopting the fast addition usually needs non-trivial memory-specific adjustments, including modifying address decoders, sense amplifiers, and other peripheral circuits. Considering the implementation overhead, the fast addition is more suitable for array-based memories like SRAM than crossbar-based memories like ReRAM.} 

\subsection{Data Mapping}
\textcolor{black}{FAT has a new data mapping scenario, where the activations are mapped to the memory arrays, and the weights are mapped to the memory controller to control the sparse dot product. Suitable data mapping brings higher performance and better energy efficiency. However, existing mapping methods for IMC platforms cannot directly apply to FAT. \cite{ISCAS_2019_PIM-Mapping} targets ReRAM crossbars where weights are stored in the memory cells and activations are mapped to Word-Lines. \cite{TVLSI_2017_STT-CiM} maps both activations and weights to the memory arrays. FAT cannot adopt these two mapping schemes due to architecture differences. BWN accelerators ParaPIM \cite{ASPDAC_2019_ParaPIM} and MRIMA \cite{TCAD_2019_MRIMA} only provide the mapping of computing one output feature map point, which does not apply to FAT due to the lack of high-level data movement.}

\textcolor{black}{Therefore, we review the data mapping schemes and present the Combined-Stationary mapping to optimize memory utilization and parallelism, taking the convolution layer as an example.}
\begin{figure}[!b] 
\centering 
\includegraphics[width=0.48\textwidth]{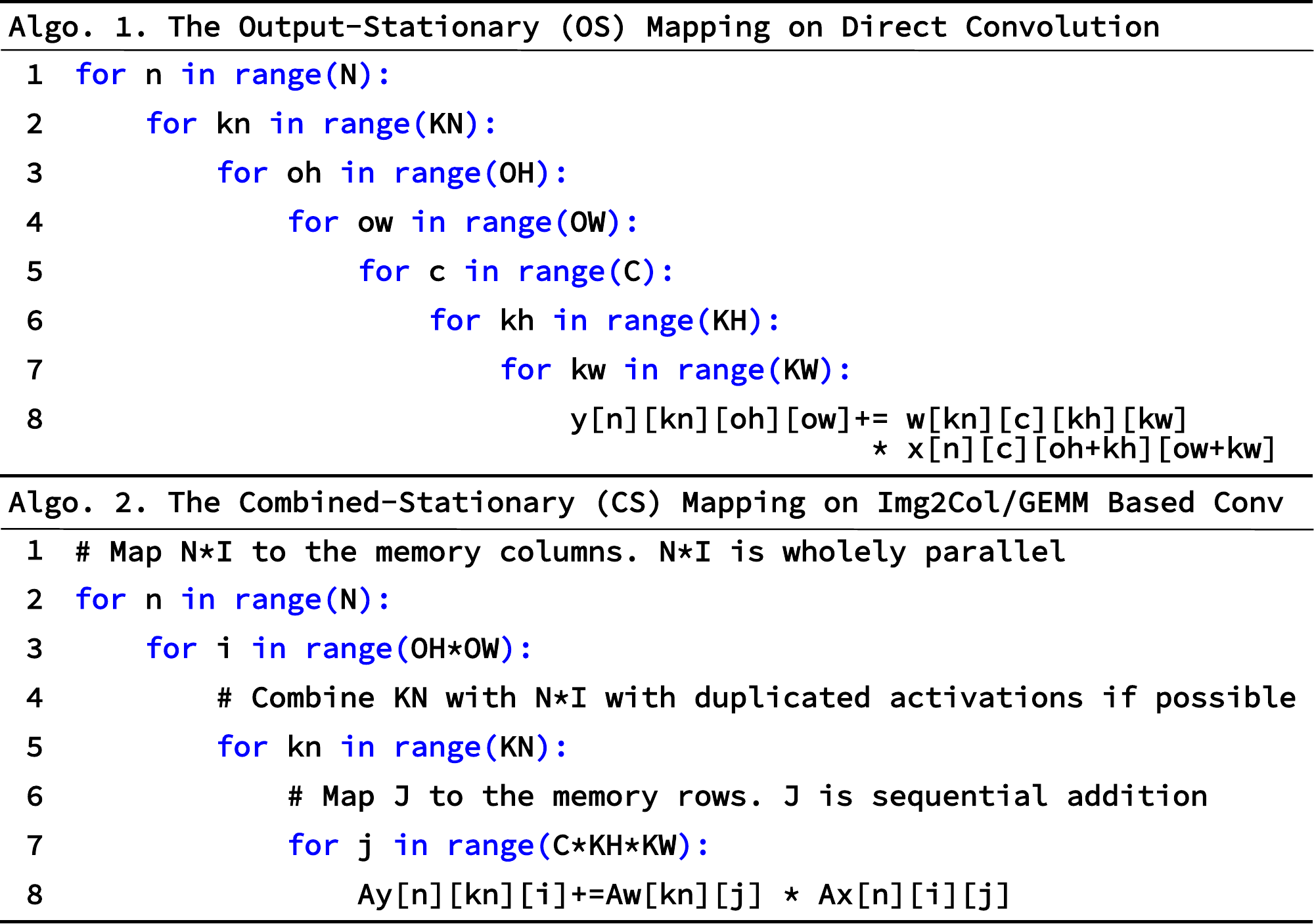} 
\caption{\textcolor{black}{The Output-Stationary and Combined-Stationary mappings.}} 
\label{Fig-map-algo}
\end{figure}

\subsubsection{Existing Convolution and Mapping Methods}
\textcolor{black}{Convolution is usually implemented by direct convolution or Image-to-Column (Img2Col) + General Matrix Multiplication (GEMM). In direct convolution, the weights slide across the activations and perform vector dot product, as the first algorithm in Fig. \ref{Fig-map-algo} presents.} \textcolor{black}{For an activation tensor a in shape [N, C, H, W (Batch Size, Channel, Height, Width)] and the weight tensor w in shape [KN, C, KH, KW (Number of Filters, Channel, Kernel Height, Kernel Width)], the shape of output feature map tensor y is [N, OC, OH, OW (Batch Size, Output Channel, Output Height, Output Width)]. As the output channel OC equals the filter number KN, we use kn to access the second dimension of y in line 8 of Algo.1.} Direct convolution is straightforward to implement \textcolor{black}{and} can reduce memory usage due to the data reuse across sliding windows. However, \textcolor{black}{the convolution stride S can decrease its data reuse.} Furthermore, direct convolution is a sequential algorithm inefficient for achieving high parallelism.

\begin{figure}[!tb] 
\centering 
\color{black}
\includegraphics[width=0.48\textwidth]{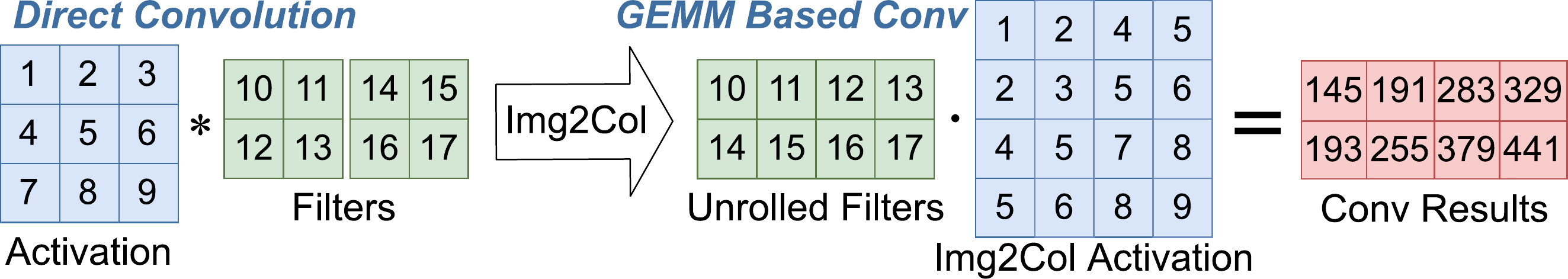} 
\caption{Img2Col transforms convolution into GEMM.} 
\label{Fig-map-img2col}
\end{figure}

Therefore, Img2Col/GEMM based convolution is introduced to reduce the data dependency and improve the parallelism. \textcolor{black}{First,} Img2Col transforms the activations into 2D arrays taking the convolution stride into account, as Fig. \ref{Fig-map-img2col} shows. Then a GEMM between the unrolled weights and the transformed activations computes the convolution results. Img2Col only keeps the activations needed for convolution \textcolor{black}{to maximize the data-level parallelism.} 

There are four main ways of mapping the convolution data \textcolor{black}{to the hardware for execution,} namely Weight-Stationary (WS), Input-Stationary (IS), Output-Stationary (OS) and Row-Stationary (RS) \cite{JCS_2019_Eyeriss-v2, CAN_2016_Eyeriss, MICRO_2019_DataFlow}. WS is optimized to reuse the weights and reduce the movement of weights. Similarly, IS and OS are optimized to reuse the activations and the (intermediate) convolution results, respectively. While RS \textcolor{black}{optimizes DRAMs' memory access,} accessing the DRAM data in consecutive rows has less latency and energy. Fig. \ref{Fig-map-algo} provides an example of OS mapping \textcolor{black}{that} keeps the output feature map tensor y in the inner loop unchanged as long as possible. 


\begin{figure}[!b] 
\centering 
\color{black}
\includegraphics[width=0.485\textwidth]{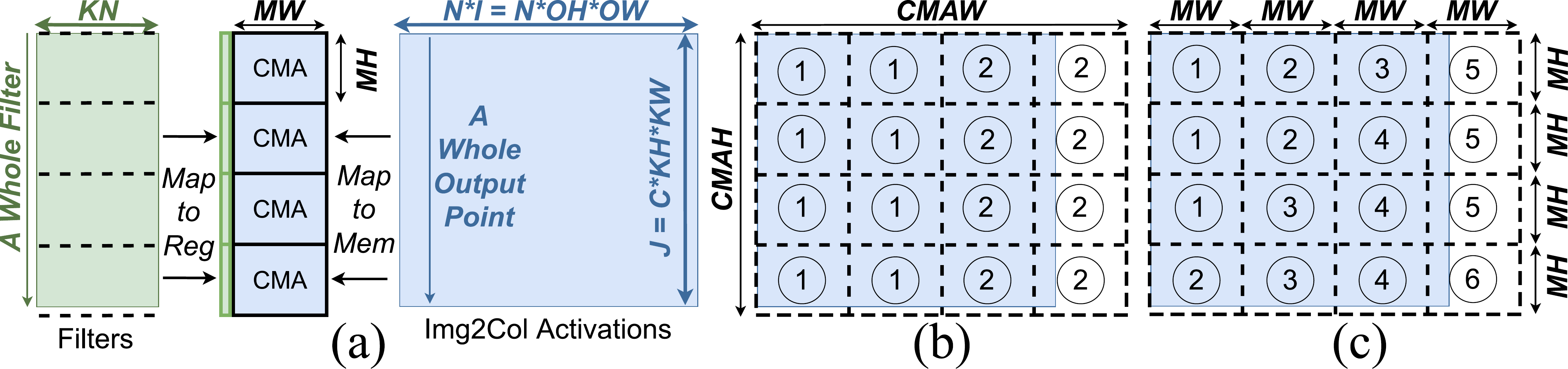} 
\caption{(a) Mapping of filters and activations to the Computing Memory Arrays (CMAs). (b) Computation sequence on the example activation matrix with eight CMAs. (c) Computation sequence with three CMAs.} 
\label{Fig-fat-img2col}
\end{figure}
\begin{table*}[]
\scriptsize
\centering
\caption{Comparison of mapping methods in convolution layers.} 
\label{Tab-Map-compare} 
\color{black}
\setlength\tabcolsep{3.5pt} 
\begin{tabular}{|l|l|l|l|l|l|l|l|}

\hline
\multicolumn{1}{|c|}{\multirow{2}{*}{Mapping}} & \multicolumn{2}{c|}{X/Ax{[}N{]}{[}I{]}{[}J{]} Loading}                                                & \multicolumn{2}{c|}{W/Aw{[}KN{]}{[}J{]} Loading}                                                                 & \multicolumn{1}{c|}{\multirow{2}{*}{\begin{tabular}[c]{@{}c@{}}Parallel Columns\end{tabular}}} & \multicolumn{1}{c|}{\multirow{2}{*}{Occupied CMAs}} & \multicolumn{1}{c|}{\multirow{2}{*}{Computing Time}}                               \\ \cline{2-5}
\multicolumn{1}{|c|}{}                         & \multicolumn{1}{c|}{Data/Load}  & \multicolumn{1}{c|}{Times}                                         & \multicolumn{1}{c|}{Data/Load}        & \multicolumn{1}{c|}{Times}                                               & \multicolumn{1}{c|}{}                                                                             & \multicolumn{1}{c|}{}                               & \multicolumn{1}{c|}{}                                                              \\ \hline

Direct-OS      & \multicolumn{1}{l|}{KN*N*MH*MW} & \begin{tabular}[c]{@{}l@{}}{[}C/MH{]}*{[}H*W/MW{]}\end{tabular} & \multicolumn{1}{l|}{KN*N*MH}          & \begin{tabular}[c]{@{}l@{}}{[}C/MH{]}*KH*\\ {[}H*W/MW{]}*KW\end{tabular} & \begin{tabular}[c]{@{}l@{}}min(MW/S, H*W/S)\end{tabular}                                       & KN*N                                                & \begin{tabular}[c]{@{}l@{}}{[}C/MH{]}*{[}H*W/MW{]}\\ *KH*KW*(MH+C/MH)\end{tabular} \\ \hline
Img2Col-OS     & \multicolumn{1}{l|}{KN*N*MH*MW} & {[}J/MH{]}*{[}I/MW{]}                                              & \multicolumn{1}{l|}{KN*N*MH}          & {[}J/MH{]}*{[}I/MW{]}                                                    & min(MW, I)                                                                                        & KN*N                                                & {[}J/MH{]}*{[}I/MW{]}*(MH+J/MH)                                                    \\ \hline
Img2Col-IS     & \multicolumn{1}{l|}{N*I*J}      & 1                                                                  & \multicolumn{1}{l|}{{[}N*I/MW{]}*J}   & KN                                                                       & min(MW, N*I)                                                                                      & {[}J/MH{]}*{[}N*I/MW{]}                             & KN*(MH+J/MH)                                                                       \\ \hline
Img2Co-WS     & \multicolumn{1}{l|}{KN*J*MW}    & N*{[}I/MW{]}                                                       & \multicolumn{1}{l|}{KN*J}             & 1                                                                        & min(MW, I)                                                                                        & {[}J/MH{]}*KN                                       & N*{[}I/MW{]}*(MH+J/MH)                                                             \\ \hline
\textbf{Img2Col-CS}   & \multicolumn{1}{l|}{\textbf{L*N*I*J}}    & \textbf{1}                                                                  & \multicolumn{1}{l|}{\textbf{L*{[}N*I/MW{]}*J}} & \textbf{KN/L}                                                                     & \textbf{min(MW, N*I)}                                                                                      & \textbf{{[}2J/MH{]}*{[}N*I/MW{]}*L}                          & \textbf{KN*(MH/2+2J/MH)/L}                                                                  \\ \hline
\end{tabular}

\end{table*}

\subsubsection{Img2Col Based Combined-Stationary Mapping}

\textcolor{black}{We propose the mapping method for FAT based on the following considerations.} \underline{First}, the memory arrays are throughput driven with high parallelism across columns. Img2Col based convolution is favoured because it decouples the sliding window style direct convolution into a parallel GEMM problem. \underline{Second}, \textcolor{black}{The 2-bit weights of TWNs are loaded to the SRAM registers inside the memory controller, while the 8-bit activations are loaded to the STT-MRAM memory array. Reloading the 2-bit weights is faster and more energy-efficient than reloading the 8-bit activations. \underline{Third}, SRAM has almost unlimited write endurance, but STT-MRAM has \(\sim 10^{15}\) write operations in its lifecycle. Therefore, reusing the activations brings more memory cell endurance and write energy benefits than reusing the weights,} and keeping the activations in the memory arrays rather than moving them is essential.

\textcolor{black}{Therefore, we propose a Combined-Stationary (CS) style mapping for FAT upon the Img2Col/GEMM based convolution and Input-Stationary (IS) mapping scheme, bringing high data-level parallelism and reducing the data movement of activations. Fig. \ref{Fig-map-algo} and Fig. \ref{Fig-fat-img2col} present the algorithm and graphic illustrations of CS mapping. The Ax[N][KN][I], Aw[KN][J] and Ay[N][I][J]} in the second algorithm of Fig. \ref{Fig-map-algo} refer to the activation array, the weight array and the output feature map array after Img2Col. \textcolor{black}{The output feature maps' height and width are combined as one dimension I (I=OH*OW), which} is mapped to the memory columns horizontally for parallel processing. \textcolor{black}{The filters' channel, height, and width are combined as one dimension J (J=C*KH*KW), which} is mapped to the memory rows vertically for sequential addition. \textcolor{black}{The 2-bit zero and non-zero weights are loaded to the memory controllers of corresponding memory arrays.} 

\textcolor{black}{The CS mapping contains the following optimizations upon the IS mapping. \underline{First}, we distribute J=C*KH*KW across memory arrays in aligned columns. As Fig. \ref{Fig-fat-img2col} (a) shows, J is the vector length in the sparse dot product that caters to a whole output point. One CMA with 512 rows and 256 columns can store MH*MW=64*256 8-bit operands. As the CMA height MH=64 is usually smaller than J, storing the whole dimension J inside one CMA reduces the column parallelism by J/MH times, which also needs to load one filter in J/MH parts sequentially. Thus distributing the dimension J across J/MH CMAs can achieve parallelism across all memory columns and loads one filter to J/MH arrays once.} \underline{Second}, we store the activation in the memory arrays with intervals whose height equals one operand. As the accumulation of immediate results needs many write operations to the memory cells, \textcolor{black}{storing the immediate results in constant rows will make the memory cells in these rows reach their lifetime soon. Therefore, we store the immediate results in reserved intervals} to distribute the memory writes to half of the memory array. As a result, the memory cells' lifetime will be more balanced. \textcolor{black}{\underline{Third}, we adopt a grid-based flexible processing sequence. As Fig. \ref{Fig-fat-img2col} (a) shows, the entire Img2Col activation array to be processed contains N*I columns and J rows. We divide the whole activation matrix into sub-arrays according to the Computing Memory Array (CMA) size, then assign the activation sub-arrays to CMAs. The CMAs compute on the activation sub-arrays one by one when the activation size exceeds the total CMA capacity. We prioritize the J dimension to reuse the immediate accumulation results. For example, the activation matrix in Fig. \ref{Fig-map-img2col} (a) can be divided into 16 CMA sizes. We map the activation sub-arrays to the CMAs in two steps when eight CMAs are available, as Fig. \ref{Fig-map-img2col} (b) shows. We load the activation sub-arrays once and compute on all corresponding weights to avoid activation reloading. Similarly, the activations can be mapped to three CMAs in six steps, as Fig. \ref{Fig-map-img2col} (c) shows.} The basic CS style mapping can \textcolor{black}{scale up with available space.} We can duplicate the activation arrays and load more filters for higher parallelism across KN filters. 


\begin{table*}[]
\centering
\caption{Comparison of mapping methods in an example convolution layer (layer 10 of ResNet-18).} 
\label{Tab-Map-ResNet} 
\color{black}
\setlength\tabcolsep{4pt} 
\begin{tabular}{|l|r|rr|rr|rrrrrr|r|}
\hline
\multicolumn{1}{|c|}{\multirow{2}{*}{Mapping}} & \multicolumn{1}{c|}{\multirow{2}{*}{CMAs}} & \multicolumn{2}{c|}{X/Ax Loading}                            & \multicolumn{2}{c|}{W/Aw Loading}                            & \multicolumn{6}{c|}{Layer Level Performance Including Loading and Computing}                                                                                                                                                                & \multicolumn{1}{c|}{\multirow{2}{*}{\begin{tabular}[c]{@{}c@{}}Max Single \\ Cell Write\end{tabular}}} \\ \cline{3-12}
\multicolumn{1}{|c|}{}                         & \multicolumn{1}{c|}{}                      & \multicolumn{1}{c|}{Time (ns)} & \multicolumn{1}{c|}{Writes} & \multicolumn{1}{c|}{Time (ns)} & \multicolumn{1}{c|}{Writes} & \multicolumn{1}{c|}{Para. Cols} & \multicolumn{1}{c|}{Utilization} & \multicolumn{1}{c|}{Time (ns)} & \multicolumn{1}{c|}{Speedup} & \multicolumn{1}{c|}{Energy (J)} & \multicolumn{1}{c|}{E. Ratio} & \multicolumn{1}{c|}{}                                                                       \\ \hline
Direct-OS     & 4096      & \multicolumn{1}{r|}{21668}     & 3.29M     & \multicolumn{1}{r|}{12437}     & 0.59K   & \multicolumn{1}{r|}{128/256} & \multicolumn{1}{r|}{76.56\%} & \multicolumn{1}{r|}{71314}     & \multicolumn{1}{r|}{1.00\(\times\)}   & \multicolumn{1}{r|}{4.295}      & 100.0\%                       & 64\(\times\)                                                                                          \\ \hline
Img2Col-OS    & 4096      & \multicolumn{1}{r|}{48753}     & 7.40M     & \multicolumn{1}{r|}{3105}      & 1.34K   & \multicolumn{1}{r|}{196/256} & \multicolumn{1}{r|}{76.56\%} & \multicolumn{1}{r|}{60883}     & \multicolumn{1}{r|}{1.17\(\times\)}   & \multicolumn{1}{r|}{7.058}      & 164.3\%                       & 64\(\times\)                                                                                          \\ \hline
Img2Col-IS   & 4096      & \multicolumn{1}{r|}{2708}      & 0.51M      & \multicolumn{1}{r|}{2523}      & 1.09K    & \multicolumn{1}{r|}{256/256} & \multicolumn{1}{r|}{94.23\%} & \multicolumn{1}{r|}{14622}     & \multicolumn{1}{r|}{4.88\(\times\)}   & \multicolumn{1}{r|}{2.440}      & 56.8\%                        & 64\(\times\)                                                                                          \\ \hline
Img2Co-WS   & 4096       & \multicolumn{1}{r|}{48753}     & 7.40M      & \multicolumn{1}{r|}{169}       & 0.08K   & \multicolumn{1}{r|}{196/256} & \multicolumn{1}{r|}{76.56\%} & \multicolumn{1}{r|}{60481}     & \multicolumn{1}{r|}{1.18\(\times\)}   & \multicolumn{1}{r|}{7.057}      & 164.3\%                       & 64\(\times\)                                                                                          \\ \hline
\textbf{Img2Col-CS}    & \textbf{4096}     & \multicolumn{1}{r|}{\textbf{1354}}      & \textbf{0.51M}      & \multicolumn{1}{r|}{\textbf{1259}}      & \textbf{1.09K}    & \multicolumn{1}{r|}{\textbf{256/256}} & \multicolumn{1}{r|}{\textbf{47.11\%}} & \multicolumn{1}{r|}{\textbf{10400}}     & \multicolumn{1}{r|}{\textbf{6.86\(\times\)}}   & \multicolumn{1}{r|}{\textbf{2.449}}      & \textbf{57.0\%}                        & \textbf{1\(\times\)}                                                                                           \\ \hline
\end{tabular}

\end{table*}

\subsubsection{Mapping Method Analysis}
We compare the mapping methods in Table \ref{Tab-Map-compare} and Table \ref{Tab-Map-ResNet}. The compared mapping methods include the OS style mapping in direct convolution (Direct-OS), the OS, \textcolor{black}{IS, WS and our proposed CS style mapping in Img2Col/GEMM based convolution (Img2Col-OS/WS/IS/CS).} \textcolor{black}{We implement these mapping methods on FAT with optimized parallelism and data reuse.} 

Table \ref{Tab-Map-compare} shows the data loading of activation tensor \textcolor{black}{X/Ax and filter tensor W/Aw in the first four columns.} The MW and MH stand for the memory width and height, e.g., how many operands \textcolor{black}{one memory column can store.} The S stands for the convolution stride. \textcolor{black}{As the memory arrays load data in parallel and most memory arrays conduct full loading in these mapping schemes, the load times reflects the loading time linearly. Meanwhile, the data loading energy can be estimated using the total data load amount by multiplying the data/load and the load times.} Img2Col based convolution improves the parallelism by dealing with the stride in the Img2Col transformation. Img2Col-WS reduces the data loading on weights, and Img2Col-IS reduces the data loading on activations compared with Img2Col-OS, which correspond to their features. \textcolor{black}{Our Img2Col-CS based on IS keeps the high upper bound of parallel columns and makes a trade-off between the loading of activations and weights for higher performance, where L is the unrolling factor across KN. The execution time in Table \ref{Tab-Map-compare} corresponds to the number of occupied memory arrays in perfect conditions without scaling-up or scaling-down. Thus we need to compare their execution time on the same convolution layer and IMC device.}

We take layer 10 of ResNet-18 as an example to give a showcase of the actual performance of these mapping methods in Table \ref{Tab-Map-ResNet}, where (N, C, H, W)=(5, 128, 28, 28), (KN, KH, KW)=(256, 3, 3), stride S=2, \textcolor{black}{and (MH, MW)=(64, 256). The time and energy are calculated referring to \cite{GLVLSI_2020_Bench-CiM}. We scale up the mapping methods to use the same 4096} CMAs for fair comparison, as more CMAs provide higher performance. 

\textcolor{black}{\underline{First}, Img2Col-IS mapping achieves the fewest memory write on activations, the highest 256 parallel columns (Para. Cols), and the highest memory utilization (94.23\%) because it unrolls the activation across N*I to fill more CMAs and reuses activations. 
\underline{Second}, Img2Col-WS mapping achieves the shortest weight loading time and fewest write times as it tries to keep the weights unmoved.
\underline{Third}, Img2Col-OS and Img2Col-WS have small speedup values due to long activation loading time than Direct-OS. As one CMA only need one column of 2-bit weights, the weight loading time and memory writes are much smaller than the activations. Thus reusing the weights reduces less loading time than reusing the activations. \underline{Fourth}, our Img2Col-CS mapping keeps the same memory writes on activations and weights as Img2Col-IS mapping. The reserved intervals inside the CMAs reduce the effective MH from 64 to 32. Thus our Img2Col-CS mapping has shorter activation and weight loading time than Img2Col-IS. Smaller MH also reduces the sequential additions at each CMA but increases the immediate results 2\(\times\). As a result, Img2Col-CS achieves the highest 6.86\(\times\) speedup with 0.2\% slightly higher energy than Img2Col-IS. 
\underline{Fifth}, Img2Col-CS balances the maximum cell write by 64\(\times\) (or up to MH\(\times\)) compared with other mappings at the cost of half memory utilization of activations. The other half of the memory rows store immediate addition results, which significantly increases the lifetime of the CMAs.}

\section{Evaluation}
We implement the Sense Amplifiers (SAs) of FAT and related works using NCSU 45nm FreePDK45\texttrademark  \(\ \)library \cite{FreePDK45,ICMSE_2009_FreePDK}. We adopt the comparator of STT-CiM \cite{TVLSI_2017_STT-CiM} and include the output selector for all the four SAs for a fair comparison. We build, verify and evaluate the circuits in Cadence Virtuoso IC6.1.8 using Virtuoso ADEL, Spectre and Layout Suite XL to obtain the corresponding latency, power and area. We refer to \cite{MWSCAS_2017_STT-MRAM} for the write time of the 1-Transistor-1-Junction STT-MRAM memory array implemented in the same 45nm process.

The performance gain of our accelerator comes from the improvements in the addition operation and the sparsity. \textcolor{black}{Keeping the standard STT-MRAM array which accounts for around 85\% area in a CMA \cite{TVLSI_2017_STT-CiM} unchanged, we only modify the SA and add a SACU to the memory controller.} The first subsection evaluates the SAs on the performance and efficiency of IMC operations. Then we evaluate the network level performance with sparsity in the second subsection.

\subsection{Sense Amplifier Level Performance}

\subsubsection{Latency and Power of In-Memory-Computing Operations}
\begin{figure}[!tb] 
\centering 
\includegraphics[width=0.48\textwidth]{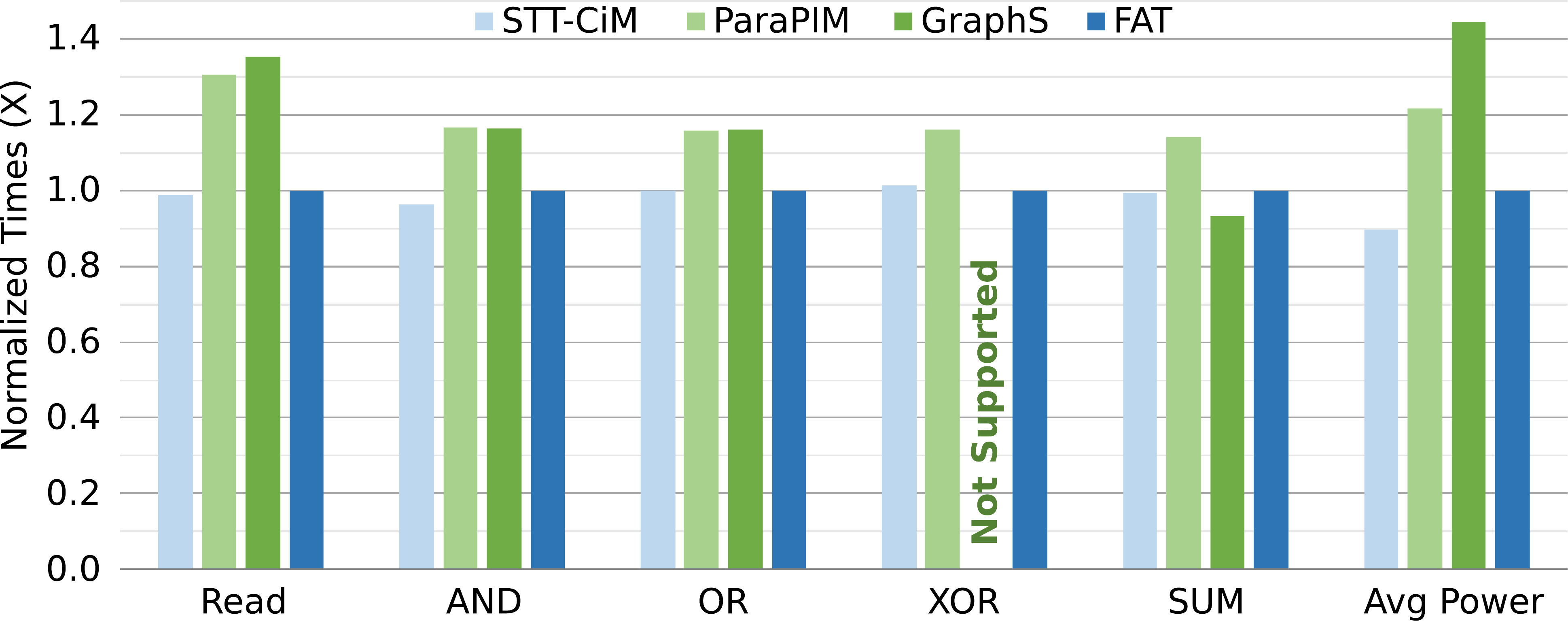} 
\caption{\textcolor{black}{Normalized critical path latency and average dynamic power of Sense Amplifiers of related works and our FAT (baseline) on IMC operations.}}
\label{Fig-SA-Latency} 
\end{figure}

We first compare the performance of IMC operations in different SAs, including STT-CiM \cite{TVLSI_2017_STT-CiM}, ParaPIM \cite{ASPDAC_2019_ParaPIM}, GraphS \cite{DATE_2019_GraphS} and our proposed FAT. Fig.\ref{Fig-SA-Latency} presents \textcolor{black}{the normalized latency and normalized average dynamic power} of the Read, AND, OR, XOR, and SUM operations. The SA latency calculates from receiving the sensing signal coming from the memory cells to getting the operation output at the OUT port. STT-CiM \cite{TVLSI_2017_STT-CiM} has slightly lower latency than FAT in Read (\textless1.3\%), AND (\textless3.7\%), OR (\textless0.2\%) and SUM (\textless0.7\%) due to its simple SA architecture. \textcolor{black}{However, STT-CiM has a} 1.4\% higher latency than FAT in XOR operation because our FAT has fewer loading \textcolor{black}{logic gates} at the XOR result port. Also, our FAT has four fewer configuration signals than STT-CiM. \textcolor{black}{Virtuoso produces consistent output signal graphs given the same input and configurations. Thus the small difference between STT-CiM and FAT is not caused by measurement noise but due to their similar performance.}

Our FAT outperforms ParaPIM \cite{ASPDAC_2019_ParaPIM} \textcolor{black}{by} about 30\% on Read, \textgreater15\% on AND, OR and XOR and 14\% on SUM. Our reusable output port design only needs a 4-to-1 selector rather than an 8-to-1 selector, which contributes to less latency of FAT. Our FAT is 35\% faster than GraphS \cite{DATE_2019_GraphS} on Read, and \textgreater15\% faster on AND and OR. GraphS is 7\% faster than FAT on SUM thanks to its aggressive computation scheme, but it does not support XOR. Our FAT is also 1.22\(\times\) and 1.44\(\times\) more power-efficient than ParaPIM and GraphS, respectively. 

\begin{table}[t]
\centering
\caption{The Critical-Path (CP) and latency of addition operations.} 
\label{Tab-ADD-8-bit}
\begin{tabular}{|c|r|r|r|r|r|r|}
\hline
Type       & \multicolumn{2}{c|}{Scalar ADD}                            & \multicolumn{4}{c|}{Vector ADD}                                                                                     \\ \hline
Bitwidth  & \multicolumn{2}{c|}{8-bit}                             & \multicolumn{2}{c|}{8-bit}                             & \multicolumn{2}{c|}{16-bit}                            \\ \hline
Time (ns) & \multicolumn{1}{c|}{CP} & \multicolumn{1}{c|}{Latency} & \multicolumn{1}{c|}{CP} & \multicolumn{1}{c|}{Latency} & \multicolumn{1}{c|}{CP} & \multicolumn{1}{c|}{Latency} \\ \hline 
STT-CiM \cite{TVLSI_2017_STT-CiM}   & \textbf{0.41}           & \textbf{8.91}                & 3.26                    & 71.26                        & 10.85                   & 146.85                       \\ \hline
ParaPIM \cite{ASPDAC_2019_ParaPIM}  & 2.47                    & 138.47                       & 2.47                    & 138.47                       & 4.95                    & 276.95                       \\ \hline
GraphS \cite{DATE_2019_GraphS}   & 1.18                    & 137.18                       & 1.18                    & 137.18                       & 2.36                    & 274.36                       \\ \hline
\color{black}
\textbf{FAT (ours)}       & \color{black} \textbf{1.13}           & \color{black} \textbf{69.13}               & \color{black} \textbf{1.13}           & \color{black} \textbf{69.13}               & \color{black} \textbf{2.26}           & \color{black} \textbf{138.26}              \\ \hline
\end{tabular}
\end{table}

\subsubsection{Latency and Efficiency of Addition Operation}
We present the Critical Path (CP) in the SA and the latency of addition operations in related works and our FAT in Table \ref{Tab-ADD-8-bit}. We include the time of writing back the result to the memory array in the latency of 8-bit scalar addition and the 8-bit /16-bit vector addition. The addition schemes of STT-CiM, ParaPIM, GraphS and our proposed FAT are different. STT-CiM computes the 8-bit addition in one step, as mentioned in Fig. \ref{Fig-add}, \textcolor{black}{while} the other three methods perform the 8-bit addition in 8 steps. \textcolor{black}{Thus} STT-CiM has the shortest critical path and latency when performing one scalar addition. The ParaPIM series, GraphS series, and proposed FAT style addition conduct the addition bit by bit. One 8-bit addition and adding two 8-bit vectors have the same eight steps in the bit-by-bit style addition, as long as the vector length does not exceed the memory array width. \textcolor{black}{Thus} adding two scalars and two vectors share the same latency in these three methods. Our FAT outperforms ParaPIM and GraphS in the critical path, single addition and vector addition thanks to the new SA and the proposed efficient addition scheme, which stores the carry in a latch instead of the memory array. FAT has shorter latency than STT-CiM on vector addition because STT-CiM has to repeat the addition for N times in the N-bit vector addition. 

\begin{figure}[!b] 
\centering 
\includegraphics[width=0.485\textwidth]{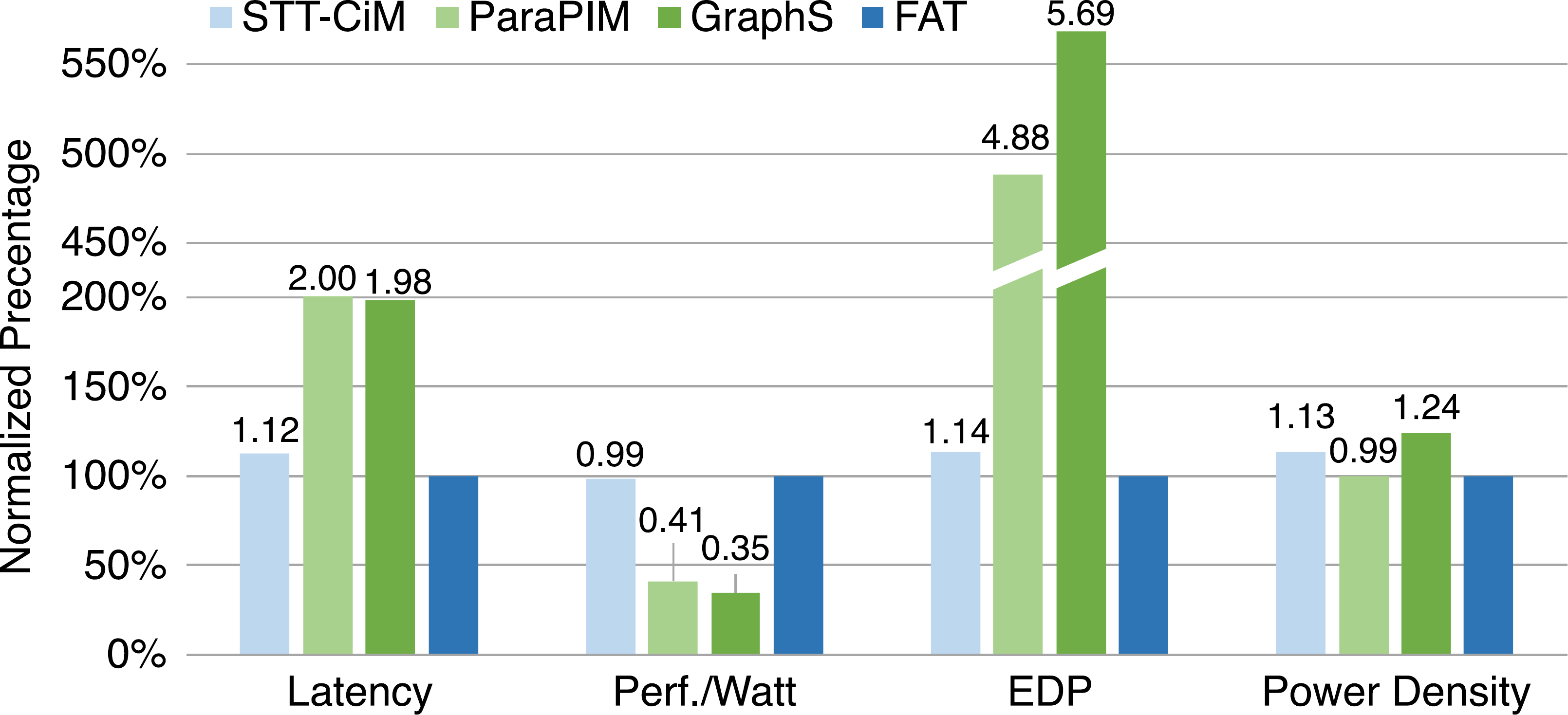} 
\caption{\textcolor{black}{Normalized latency and efficiency of 32-bit Addition of our proposed FAT and related works (Baseline: FAT).}} 
\label{Fig-32-bit-ADD} 
\end{figure}

Our FAT builds the SA for the most efficient vector addition among related works. As neural networks are throughput-driven and TWNs utilize vector additions in convolution layers, vector addition is more important than single scalar addition in TWN accelerators. \textcolor{black}{We take the 32-bit vector addition to analyze further the performance and efficiency of these SAs in Fig. \ref{Fig-32-bit-ADD}, including the latency, performance/watt, Energy-Delay-Product (EDP), and power density (power/area).}  

\underline{First}, FAT is 1.12\(\times\), 2.00\(\times\) and 1.98\(\times\) faster than STT-CiM, ParaPIM and GraphS respectively in 32-bit vector addition. Taking the write overhead of the SUM and Carry into account, we find that ParaPIM and GraphS are much slower than STT-CiM due to the extra write of Carry. Though GraphS has less computation latency in the SA, \textcolor{black}{writing the Carry to memory is so time-consuming that it slows down the whole vector addition. Our FAT avoids the propagation of Carry and has a shorter SUM critical path than STT-CiM. FAT also avoids the extra write of Carry in ParaPIM and GraphS. Thus FAT has the shortest vector addition latency.} \underline{Second}, FAT has the highest performance/watt and is 1.01-2.86\(\times\) as efficient \textcolor{black}{as} related works. \textcolor{black}{\underline{Third}, our FAT has the least EDP among related works and is 1.14-5.69\(\times\) as efficient as STT-CiM, ParaPIM and GraphS.} \underline{Fourth}, FAT has a lower power density than STT-CiM and GraphS, which means our FAT is more balanced on the power distribution and may have a longer lifetime. 

In summary, we realize a vector addition with high performance and high efficiency for TWNs by innovation from the addition scheme and the SA circuit.  

\subsubsection{Area and Reliability of Sense Amplifiers}
\begin{figure}[!tb] 
\centering 
\includegraphics[width=0.48\textwidth]{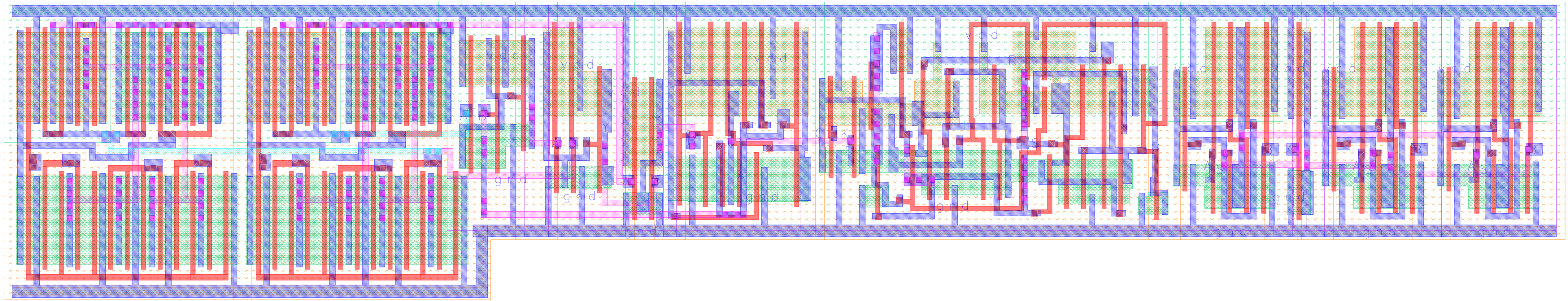} 
\caption{Layout of our Sense Amplifier in Virtuoso Layout Suite XL.} 
\label{Fig-Layout} 
\end{figure}

\begin{figure}[!b] 
\centering 
\includegraphics[width=0.36\textwidth]{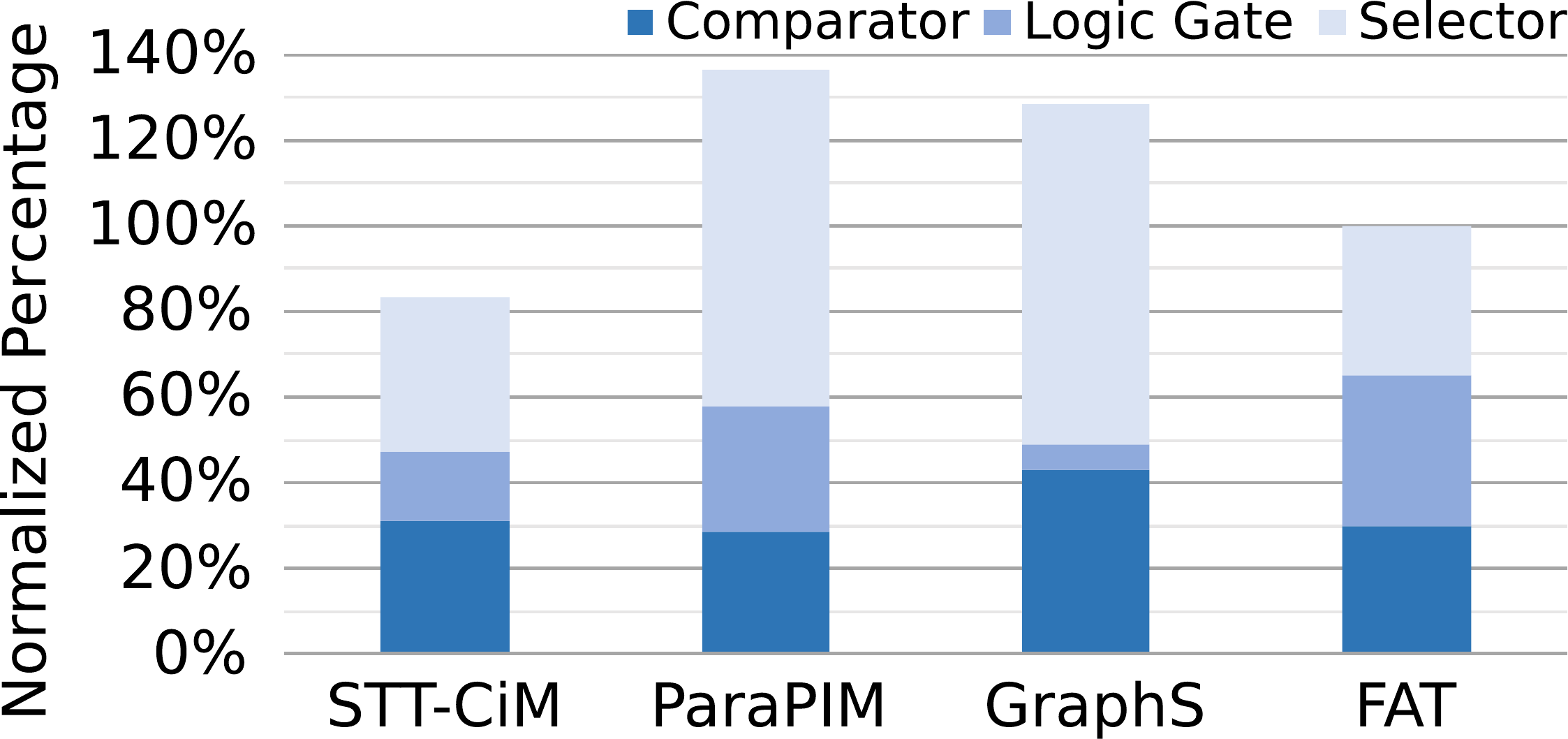} 
\caption{\textcolor{black}{Normalized area breakdown of the Sense Amplifiers in our proposed FAT and related works (Baseline: FAT).}} 
\label{Fig-area} 
\end{figure}

\textcolor{black}{We present the high-resolution layout figure (6800x1300 pixels) of our SA in Fig.\ref{Fig-Layout} for reference, which can be zoomed in for more details.} The area of our proposed FAT and related works are shown in Fig.\ref{Fig-area}. We normalize the area to FAT for easy comparison. FAT has 21\% more area than STT-CiM because of the D-Latch, but it is 1.22\(\times\) and 1.17\(\times\) area efficient than ParaPIM and GraphS. ParaPIM has seven output ports \textcolor{black}{requiring an 8-input selector and a D-Latch. Thus ParaPIM has a larger size than STT-CiM with a 4-input selector and no latch. GraphS has fewer logic gates but one more comparator than ParaPIM, resulting in} a similar area as ParaPIM. Though FAT has the most logic gates, FAT has smaller area than ParaPIM and GraphS due to only two comparators and a 4-input selector. 

Also, our SA is more reliable than ParaPIM and GraphS. FAT's SA only contains two-operand operations rather than three-operand operations like ParaPIM and GraphS. The sense margin of two-operand operations is 2.4\(\times\) as high as that of three-operand operations \cite{ASPDAC_2019_ParaPIM, TCAD_2019_MRIMA, DATE_2019_GraphS, NanoArc_2019_ParaPIM}. As a larger sense margin brings less error rate, our SA is more reliable than ParaPIM and GraphS. 

\subsection{Network Level Performance}
\begin{figure}[!tb] 
\centering 
\includegraphics[width=0.48\textwidth]{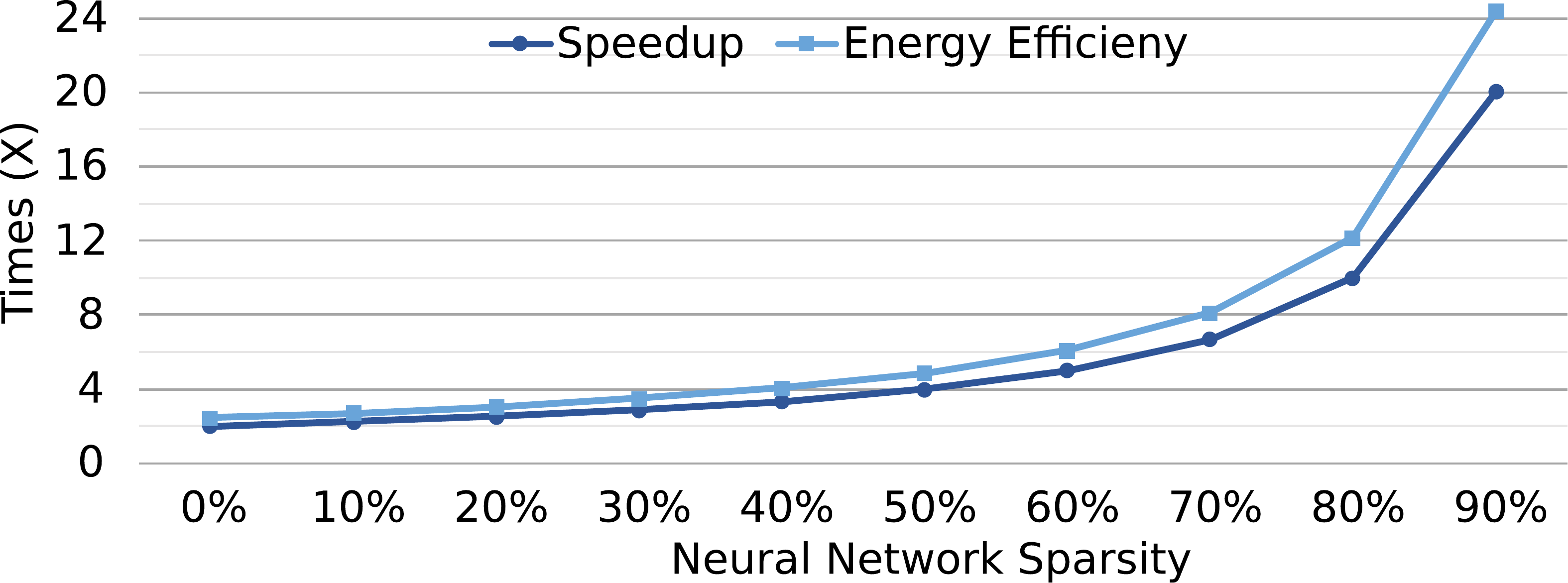} 
\caption{\textcolor{black}{Speedup and energy efficiency of our FAT across network sparsity (Baseline: ParaPIM \cite{ASPDAC_2019_ParaPIM}).}} 
\label{Fig-speedup-sparsity} 
\end{figure}

\textcolor{black}{Fig. \ref{Fig-speedup-sparsity} presents the network level speedup and energy efficiency on models with similar sparsity across layers, which also reflects the single layer speedup and efficiency.} We take ParaPIM as the baseline \textcolor{black}{because} only ParaPIM series \textcolor{black}{designs} are BWN IMC accelerators among the three series of related works. The overall performance improvement is determined by both the fast addition and the sparsity. We achieve 2.00\(\times\) speedup and 1.22\(\times\) power efficiency compared with ParaPIM from the \textcolor{black}{proposed addition operator.} The SACU further brings speedups from the sparsity of the TWNs. Our accelerator can achieve \textcolor{black}{3.34\(\times\), 5.01\(\times\), and 10.02\(\times\) network level speedup and 4.06\(\times\), 6.09\(\times\), and 12.19\(\times\) energy efficiency compared with ParaPIM when the average sparsity is 40\%, 60\%, and 80\%.} \textcolor{black}{Since our mapping in section III.C performs dense mapping, and then the proposed SACU exploits the fine-grained sparsity of the filters, the speedup and energy efficiency are independent of layer sizes and the model architectures.}


\section{Conclusion}
In this paper, we propose FAT as an In-Memory-Computing accelerator with fast addition and sparse dot product for Ternary Weight Neural Networks. We propose an efficient in-memory addition scheme and a new Sense Amplifier that stores the carry inside the Sense Amplifier to avoid the carry propagation latency and carry writing back latency. Our Sense Amplifier is 2.00\(\times\) faster, 1.22\(\times\) power-efficient and 1.22\(\times\) area efficient than ParaPIM and MRIMA on addition operations. We propose a Sparse Addition Control Unit which utilizes the sparsity of TWNs to skip the addition operations corresponding to zero weights. FAT with the sparse dot product achieves up to 10.02\(\times\) speedup and 12.19\(\times\) energy efficiency on networks with 80\% sparsity compared with ParaPIM and MRIMA. We further present the Combined-Stationary mapping to reach near 100\% column parallelism across the whole memory array \textcolor{black}{and balance the STT-MRAM cell write times for longer memory lifetime, which brings 6.86\(\times\) speedup and reduces 43.0\% energy 
compared with direct convolution based Output-Stationary mapping on ResNet-18 layer 10.}

\section*{Acknowledgment}
This work is partially supported by the Ministry of Education, Singapore, under its Academic Research Fund Tier 2 (MOE2019-T2-1-071) and Tier 1 (MOE2019-T1-001-072), and partially supported by Nanyang Technological University, Singapore, under its NAP (M4082282) and SUG (M4082087).

\( \)


\Urlmuskip=0mu plus 1mu
\bibliographystyle{IEEEtran}
\bibliography{BWNPIM}

\end{document}